\documentclass[12pt]{article}
\usepackage[latin9]{inputenc}
\usepackage{mathtools}
\usepackage{amsmath}
\usepackage{amsthm}
\usepackage{amssymb}
\usepackage[a4paper]{geometry}
\usepackage{setspace}
\usepackage[authoryear,round]{natbib}
\usepackage{tikz}
\usepackage{booktabs}
\usepackage[extra]{tipa}
\makeatletter

\providecommand{\tabularnewline}{\\}

\theoremstyle{plain}
\newtheorem{thm}{\protect\theoremname}
\theoremstyle{definition}
\newtheorem{example}[thm]{\protect\examplename}
\theoremstyle{plain}
\newtheorem{prop}[thm]{\protect\propositionname}

\renewcommand{\hat}{\widehat}
\renewcommand{\tilde}{\widetilde}

\newtheorem{assumption}{Assumption}\newtheorem{defi}{Definition}

\date{\today}

\providecommand{\examplename}{Example}
\providecommand{\propositionname}{Proposition}

\providecommand{\theoremname}{Theorem}

\makeatother

\providecommand{\examplename}{Example}
\providecommand{\propositionname}{Proposition}
\providecommand{\theoremname}{Theorem}

\begin{document}
\title{The Generalized Falsification Adaptive Set for Violations of the Exclusion
Restriction and Exogeneity}
\author{Nicolas Apfel$^{a}$ and Frank Windmeijer$^{b}$ \\
 $^{a}${\small Dept of Economics, University of Innsbruck, Austria}\\
 $^{b}${\small Dept of Statistics and Nuffield College, University
of Oxford, UK}\\
 }
\maketitle
\begin{abstract}
\noindent\baselineskip=15pt


\medskip
\noindent The falsification adaptive set (\textit{FAS}) as proposed by \citet{MastenPoirierEcta2021} provides an identified set for a treatment effect when the baseline model is falsified, assuming invalid instruments violate exclusion only. 
We show that whether an invalid instrument is a confounder or collider has important consequences: incorrect treatment can cause the \textit{FAS} to exclude the true parameter, $\beta$. We derive pattern-specific falsification adaptive sets for each combination of violations and propose a generalized \textit{FAS} as their union, containing $\beta$ if any instrument is valid. 
We illustrate our results with the roads and trade application of \citet{Duranton2014}.
\end{abstract}
\noindent{\small\textbf{Keywords:}}{\small{} Instrumental variables,
invalid instruments, falsification adaptive set}{\small\par}

 {\small\bigskip{}
 }{\small\par}

\noindent\rule{4cm}{0.4pt}\\
 \begin{small} Addresses for correspondence: 

Nicolas Apfel: nicolas.apfel@uibk.ac.at, 

Frank Windmeijer: frank.windmeijer@stats.ox.ac.uk.

Nicolas Apfel gratefully acknowledges funding through ESRC grant EST013567/1.
\end{small}

\thispagestyle{empty}

\baselineskip=20pt

\pagebreak{}

\pagenumbering{arabic} \setcounter{page}{1}

\section{Introduction}

Instrumental variables (IVs) are widely used in economics to estimate treatment effects. The key identifying assumptions are that the instruments affect the outcome only through the treatment (the exclusion restriction) and that they are uncorrelated with unobserved determinants of the outcome (exogeneity).

\citet[henceforth MP]{MastenPoirierEcta2021} introduce the falsification adaptive set (\textit{FAS}), which is the identified set for the treatment effect under the minimal relaxation of the baseline assumptions such that the model is just not falsified. Even when the researcher cannot point-identify the parameter of interest, the \textit{FAS} provides a range of values consistent with these minimal violations.

In this paper, we distinguish between different reasons for an IV being invalid and show that this has important consequences for the \textit{FAS}. An invalid instrument may act as a \textit{confounder}, violating the exclusion restriction through a direct effect on the outcome, or as a \textit{collider}, when the instrument instead is influenced by the error term. This distinction is well established in the causal inference literature. The two types of violations require opposite treatments: a confounder should be controlled for, whereas a collider should be excluded from the analysis.

We make three contributions. First, we show that if there is a valid and relevant IV, the \textit{FAS} is guaranteed to contain the true treatment effect parameter $\beta$ only if the invalid IVs are treated according to their true violation type. Incorrectly treating a violation of exclusion as a violation of exogeneity, or vice versa, can cause the \textit{FAS} to exclude $\beta$, which is misleading when reporting the \textit{FAS} as an identified set. Correct treatment of exogeneity violations leads to a different \textit{FAS} for violations of the exogeneity assumption, which we denote \textit{FAS}$_{exo}$, as opposed to the original \textit{FAS}$_{excl}$ of MP. Second, we derive a pattern-specific \textit{FAS} for each possible combination of exclusion and exogeneity violations across instruments, and propose the generalized \textit{FAS} as their union. If at least one instrument is valid and relevant, the generalized \textit{FAS} is guaranteed to contain $\beta$ without requiring the researcher to specify the violation type of each instrument ex ante. Third, we provide an alternative derivation of the falsification frontier from the population 2SLS objective function, clarifying its interpretation as a minimal relaxation leading to nonfalsification regardless of the actual violation type.

We do not prefer one \textit{FAS} over another. Our generalized \textit{FAS} is the appropriate set when the researcher is agnostic about how each invalid instrument violates the identifying assumptions. It can be sharpened to a pattern-specific \textit{FAS}, which is weakly narrower than the generalized \textit{FAS}, when domain knowledge pins down the violation type of some or all instruments.

We illustrate these results using the empirical analysis of roads and trade by \citet[DMT]{Duranton2014}, which has also been used by MP. DMT estimate the causal effect of within-city interstate highways on a city's propensity to export, using three historical instruments: planned highways from a 1947 federal map, railroad routes from 1898, and historical exploration routes. With three correlated instruments this is a natural setting to study the \textit{FAS}. Because some historical instruments might have a direct effect on current trade patterns, and others might be correlated with the outcome through correlation with unobserved regional characteristics, this application highlights the practical importance of distinguishing between violation types.

Sections \ref{sec:ModAss} sets up the model and sections \ref{sec:FASExcl} and \ref{sec:FASExo} develop \textit{FAS}$_{excl}$ and \textit{FAS}$_{exo}$. Section \ref{sec:Alternative-Derivation} gives an alternative derivation from the population 2SLS objective function. Section \ref{sec:ExoExcl} derives the generalized \textit{FAS} for arbitrary patterns of violations. Section \ref{sec:Empi} illustrates the results using the roads and trade data of \citet{Duranton2014}.

\section{Model and Assumptions}

\label{sec:ModAss}

We consider the linear model specification 
\begin{align}
Y & =X\beta+E,\label{eq:strucmod}
\end{align}
with $Y$ and $X$ observable scalar random variables, $E$ an unobservable
scalar random variable and $\beta$ an unknown constant. There is
endogeneity in the sense that $\text{cov}\left(X,E\right)\neq0$.
We have available an observable $k_{z}$-vector $\boldsymbol{Z}$
of putative instruments with individual elements denoted as $Z_{\ell}$,
$\ell=1,\ldots,k_{z}$. For ease of exposition, a constant and exogenous
variables are partialled out wlog.

The following sufficient variation assumption is maintained as in
MP.

\begin{assumption} \label{Ass:Suff} Sufficient variation: The $k_{z}\times k_{z}$
matrix $\text{var}\left(\boldsymbol{Z}\right)$ is invertible.
\end{assumption}

If the instruments are exogenous in the sense that $\text{cov}\left(\boldsymbol{Z},E\right)=\boldsymbol{0}$,
and relevant in the sense that $\text{cov}\left(\boldsymbol{Z},X\right)\neq\boldsymbol{0}$,
the baseline model, then $\beta$ is point identified and the standard
2SLS estimator on a sample of observations $\left\{ Y_{i},X_{i},\boldsymbol{Z}_{i}^{\prime}\right\} _{i=1}^{n}$
is a consistent estimator of $\beta$. Validity of all instruments
can be falsified for example by the \citet{Sargan1958} and \citet{Hansen1982}
tests for overidentifying restrictions. This is the type of falsification
of the model that was considered by MP and for which they proposed
the \textit{FAS. }When the model is falsified and hence $\text{cov}\left(\boldsymbol{Z},E\right)\neq\boldsymbol{0}$,
one needs to consider in what way invalid instruments violate this
exogeneity condition. As in MP, we consider violations of the exclusion and (conditional) exogeneity assumptions.

The first violation we consider is that an instrument is invalid because
it is an omitted explanatory variable and has an effect on the outcome
$Y$ over and above the effect of $X$ on $Y$. This is a violation
of the exclusion assumption.

\begin{defi} \textit{Exclusion Assumption.}

\noindent Let $\text{\ensuremath{E=\boldsymbol{Z}\boldsymbol{\gamma}}}+U$,
and so 
\begin{equation}
Y=X\beta+\boldsymbol{Z} \boldsymbol\gamma+U.\label{eq:strucmodgam}
\end{equation}
An instrument $Z_{\ell}$, $\ell\in\left\{ 1,\ldots,k_{z}\right\} $,
satisfies the exclusion assumption if $\gamma_{\ell}=0$. It violates
the exclusion assumption if $\gamma_{\ell}\neq0$.\end{defi}

The second violation we consider is that an instrument is invalid
because it does not satisfy the conditional exogeneity assumption
$\text{cov}\left(Z_{\ell},U\right)=0$.

\begin{defi}\textit{ Conditional Exogeneity Assumption}.

\noindent Let $\text{cov}\left(\boldsymbol{Z},U\right)=\boldsymbol{\alpha}$.
An instrument $Z_{\ell}$, $\ell\in\left\{ 1,\ldots,k_{z}\right\} $,
satisfies the conditional exogeneity assumption if $\alpha_{\ell}=0$.
It violates the conditional exogeneity assumption if $\alpha_{\ell}\neq0$.

\end{defi}

We term the exogeneity conditional, as it is conditional on the inclusion
of instruments that violate the exclusion assumption as controls in
the model.

Figure \ref{fig:dag-violations} illustrates the two types of violations. Instrument $Z_1$ is valid. Instrument $Z_2$ violates exclusion: it has a direct effect $\gamma_2$ on $Y$ and should be included as a control. Instrument $Z_3$ violates conditional exogeneity, it is a collider, as $U$ has a causal effect on $Z_3$ represented by $\alpha_3$, and $Z_3$ should be excluded from the analysis.

\begin{figure}[t]
\centering
\begin{tikzpicture}[
    every node/.style={font=\normalsize},
    obs/.style={circle, draw, minimum size=8mm, inner sep=0pt},
    unobs/.style={circle, draw, dashed, minimum size=8mm, inner sep=0pt},
    arr/.style={->, >=stealth, thick},
    violarr/.style={->, >=stealth, thick}
  ]
  %
  \node[obs] (Z2) at (-2,2) {$Z_2$};
  \node[obs] (Z1) at (-2,0) {$Z_1$};
  \node[obs] (Z3) at (-2,-2) {$Z_3$};
  \node[obs] (X) at (1.5,0) {$X$};
  \node[obs] (Y) at (5,0) {$Y$};
  \node[unobs] (U) at (1.5,-2) {$U$};
  %
  \draw[arr] (Z1) -- (X);
  \draw[arr] (Z2) -- (X);
  \draw[arr] (Z3) -- (X);
  \draw[arr] (X) -- node[above] {$\beta$} (Y);
  \draw[arr] (U) -- (X);
  \draw[arr] (U) -- (Y);
  \draw[violarr] (Z2) to[bend left=15] node[above] {$\gamma_2$} (Y);
  \draw[violarr, <-] (Z3) -- node[below] {$\alpha_3$} (U);
  \draw[<->, thick, dashed] (Z1) -- (Z2);
  \draw[<->, thick, dashed] (Z1) -- (Z3);
  \draw[<->, thick, dashed] (Z2) to[bend right=30] (Z3);
\end{tikzpicture}%

\caption{DAG with three instruments. $Z_1$ is valid ($\gamma_1 = \alpha_1 = 0$). $Z_2$ violates the exclusion assumption with direct effect $\gamma_2 \neq 0$ on $Y$. $Z_3$ violates the conditional exogeneity assumption with $\alpha_3 \neq 0$. Dashed lines indicate correlations.}
\label{fig:dag-violations}
\end{figure}

It is important to note that the sequential structure of the violations
of the general exogeneity condition and the direction of the effects
cannot be captured by projection arguments. For
example, in the linear projection $E=\boldsymbol{Z}'\boldsymbol{c}+r$,
we have that $\text{cov}\left(\boldsymbol{Z},r\right)=0$, and $\boldsymbol{c}=\boldsymbol{\gamma}+\text{var}\left(\boldsymbol{Z}\right)\boldsymbol{\alpha}$,
thus mixing the exclusion and conditional exogeneity violations. As
$\text{var}\left(\boldsymbol{Z}\right)$ is a general variance matrix,
with correlated instruments as illustrated in the empirical example
in Section \ref{sec:Empi}, one cannot disentangle $\boldsymbol{\gamma}$
and $\boldsymbol{\alpha}$ from $\boldsymbol{c}$.

We define a valid instrument as follows.

\begin{defi}

\label{Def:ValIInst} Valid Instrument: An instrument $Z_{\ell}$,
$\ell\in\left\{ 1,\ldots,k_{z}\right\} $ is valid if both the exclusion
and conditional exogeneity assumptions hold, $\gamma_{\ell}=\alpha_{\ell}=0$.

\end{defi}

Note that with correlated instruments, a relevant valid instrument
can point-identify $\beta$ only if invalid instruments are treated
according to their violations of the exclusion or conditional exogeneity
assumptions. For example, with $k_{z}=3$ instruments, if $\gamma_{\ell}\alpha_{\ell}=0$
$\forall\ell$, $\gamma_{1}\ne0$, $\alpha_{2}\neq0$, $\gamma_{3}=\alpha_{3}=0$,
then $\beta$ is identified from the model specification $Y=X\beta+Z_{1}\gamma_{1}+U$,
using $Z_{3}$ as the excluded instrument for $X$. Note that an instrument
with $\gamma_{\ell}\alpha_{\ell}\neq0$ is itself an endogenous explanatory
variable and point-identification cannot be achieved then even with
a valid instrument for $X$ present without further assumptions, as
such an endogenous invalid instrument needs to be identified and instrumented
itself.

As discussed in MP, if the distribution of $\left(Y,X,\boldsymbol{Z}'\right)$
is such that the model is falsified then this could be due to misspecification
of model (\ref{eq:strucmod}), which assumes homogeneous and linear
treatment effects, and/or instrument invalidity. As in MP, we maintain
model (\ref{eq:strucmod}) and focus on failures of the instrument
exclusion or the conditional exogeneity assumption as reasons for
falsifying the baseline model.

\section{The Falsification Adaptive Set Under Violations of the Exclusion
Assumption}

\label{sec:FASExcl}

MP derive the falsification adaptive set under the assumption that
invalid instruments can violate the exclusion assumption only:

\begin{assumption} \label{Ass:Exo}An invalid instrument $Z_{\ell}$
violates the exclusion assumption, $\gamma_{\ell}\neq0$. All instruments
satisfy the conditional exogeneity assumption, $\alpha_{j}=0$, for
$j=1,\ldots,k_{z}$. \end{assumption}

\noindent MP make the following partial exclusion assumption.

\begin{assumption} \label{Ass:partexcl} Partial exclusion: There
are constants $\delta_{\ell}\geq0$ such that $\left|\gamma_{l}\right|\leq\delta_{\ell}$
for $\ell=1,\ldots,k_{z}$.

\end{assumption}

\noindent As MP (p 1453) argue, Assumption \ref{Ass:partexcl} is
a natural way to relax the exclusion restriction. As falsification
is due to $\boldsymbol{\gamma}\neq0$, the natural relaxation is to
consider a range of values for $\boldsymbol{\gamma}$, $-\boldsymbol{\delta}\leq\text{\ensuremath{\boldsymbol{\gamma}}}\leq\boldsymbol{\delta}$,
componentwise.

Under Assumption \ref{Ass:Exo} it follows that 
\begin{equation}
\boldsymbol{\gamma}=\text{var}\left(\boldsymbol{Z}\right)^{-1}\left(\text{cov}\left(\boldsymbol{Z},Y\right)-\text{cov}\left(\boldsymbol{Z},X\right)\beta\right).\label{eq:gama0}
\end{equation}
Under Assumptions \ref{Ass:Suff}, \ref{Ass:Exo} and \ref{Ass:partexcl},
Theorem 1 in MP then states that 
\begin{align}
\mathcal{B}_{\gamma}\left(\boldsymbol{\delta}\right) & =\left\{ b\in\mathbb{R}:-\boldsymbol{\delta}\leq\text{var}\left(\boldsymbol{Z}\right)^{-1}\left(\text{cov}\left(\boldsymbol{Z},Y\right)-\text{cov}\left(\boldsymbol{Z},X\right)b\right)\leq\boldsymbol{\delta}\right\} ,\label{eq:idset}
\end{align}
is the identified set for $\beta$, where the inequalities are componentwise.

For any $\tilde{\boldsymbol{\delta}}<\boldsymbol{\delta}$ (componentwise, with strict inequality for at least one element), the model is falsified if and only if $\mathcal{B}_{\gamma}\left(\tilde{\boldsymbol{\delta}}\right)$ is empty, which is what the standard overidentification test verifies for $\tilde{\boldsymbol{\delta}}=\boldsymbol{0}$. We use the $\tilde{\boldsymbol{\delta}}$ notation to emphasise that $\tilde{\boldsymbol{\delta}}$ need not satisfy Assumption \ref{Ass:partexcl}.

MP introduce the falsification frontier (\textit{FF}) as the minimal
set of $\tilde{\boldsymbol{\delta}}$s which lead to a non-empty set
$\mathcal{B}_{\gamma}\left(\tilde{\boldsymbol{\delta}}\right)$. For
a $\tilde{\boldsymbol{\delta}}\in FF$ this means that for any other
$\bar{\boldsymbol{\delta}}<\tilde{\boldsymbol{\delta}}$, the identified
set $\mathcal{B}\left(\bar{\boldsymbol{\delta}}\right)$ is empty
and thus falsifies the model.

Let 
\begin{equation}
\boldsymbol{\pi}:=\text{var}\left(\boldsymbol{Z}\right)^{-1}\text{cov}\left(\boldsymbol{Z},X\right);\,\,\boldsymbol{\psi}:=\text{var}\left(\boldsymbol{Z}\right)^{-1}\text{cov}\left(\boldsymbol{Z},Y\right)\label{eq:pipsi}
\end{equation}
then it follows that the identified set for $\beta$ is given by 
\[
\mathcal{B}_{\gamma}\left(\boldsymbol{\delta}\right)=\left\{ b\in\mathbb{R}:-\boldsymbol{\delta}\leq\left(\boldsymbol{\psi}-\boldsymbol{\pi}b\right)\leq\boldsymbol{\delta}\right\} .
\]
As in MP, $Z_{\ell}$ is a relevant instrument if $\pi_{\ell}\neq0$.
Let $\mathcal{L}_{rel}$ denote the set of relevant instruments, 
\begin{equation}
\mathcal{L}_{rel}=\left\{ \ell\in\left\{ 1,\ldots,k_{z}\right\} :\pi_{\ell}\neq0\right\} .\label{eq:relsetexcl}
\end{equation}
For model (\ref{eq:strucmod}), under Assumptions \ref{Ass:Suff},
\ref{Ass:Exo} and \ref{Ass:partexcl}, Proposition 2 in MP specifies
the falsification frontier as the set 
\[
FF_{excl}=\left\{ \tilde{\boldsymbol{\delta}}\left(b\right)\in\mathbb{R}_{\geq0}^{k_{z}}:\tilde{\delta}_{\ell}\left(b\right)=\left|\psi_{\ell}-b\pi_{\ell}\right|,\,\ell=1,\ldots,k_{z},\,b\in\left[\min_{\ell\in\mathcal{L}_{rel}}\frac{\psi_{\ell}}{\pi_{\ell}},\max_{\ell\in\mathcal{L}_{rel}}\frac{\psi_{\ell}}{\pi_{\ell}}\right]\right\} ,
\]
where we have added the subscript ``excl'' to denote a relaxation
of the exclusion assumption only and have defined $\tilde{\boldsymbol{\delta}}$
as a function of $b$. Note that for all $\tilde{\boldsymbol{\delta}}\left(b\right)\in FF_{excl}$,
it follows that $\mathcal{B}_{\gamma}\left(\tilde{\boldsymbol{\delta}}\left(b\right)\right)=\left\{ b\right\} $.

The falsification adaptive set, denoted \textit{FAS$_{excl}$,} is
then given in Theorem 2 of MP as 
\begin{equation}
FAS_{excl}=\cup_{\tilde{\boldsymbol{\delta}}\in FF_{excl}}\mathcal{B}_{\gamma}\left(\tilde{\boldsymbol{\delta}}\right)=\left[\min_{\ell\in\mathcal{L}_{rel}}\frac{\psi_{\ell}}{\pi_{\ell}},\max_{\ell\in\mathcal{L}_{rel}}\frac{\psi_{\ell}}{\pi_{\ell}}\right].\label{eq:FASexcl}
\end{equation}
As MP point out in their Lemma 1, for $\ell\in\mathcal{L}_{rel}$,
$\frac{\psi_{\ell}}{\pi_{\ell}}\left(=\beta+\frac{\gamma_{\ell}}{\pi_{\ell}}\right)$
is the IV/2SLS estimand in the just-identified model specification
\begin{equation}
Y=X\beta_{\ell}+\boldsymbol{Z}_{\{-\ell\}}^{\prime}\boldsymbol{\gamma}_{\left\{ -\ell\right\} }+\tilde{\mathcal{U}}_{\ell},\label{eq:jidcon}
\end{equation}
where $\boldsymbol{Z}_{\{-\ell\}}=\boldsymbol{Z}\setminus\left\{ Z_{\ell}\right\} $,
and using $Z_{\ell}$ as the excluded just-identifying instrument,
see also \citet[Appendix A.5]{WindmeijeretalJRSSB2021}. As just-identified
models are not falsifiable, it follows that for $\ell\in\mathcal{L}_{rel}$,
$\tilde{\delta}_{\ell}\left(\frac{\psi_{\ell}}{\pi_{\ell}}\right)=0$.
From this, the results of the falsification frontier \textit{FF}$_{excl}$
follow straightforwardly, as when moving $b$ from $\min_{j\in\mathcal{L}_{rel}}\frac{\psi_{j}}{\pi_{j}}$
to $\max_{j\in\mathcal{L}_{rel}}\frac{\psi_{j}}{\pi_{j}}$ there is
at least one element in $\tilde{\boldsymbol{\delta}}\left(b\right)$
that decreases in value and at least one that increases in value.

MP (p. 1449) characterize the falsification frontier as the set of
smallest relaxations of the baseline model which are not falsified.
They recommend to report estimates of \textit{FAS}$_{excl}$ which
is an identified set for $\beta$\textit{ under the assumption that
the true model lies on the falsification frontier}. MP do not relate
this to the presence of valid instruments. As we can write the \textit{FAS$_{excl}$}
alternatively as 
\[
FAS_{excl}=\left[\beta+\min_{\ell\in\mathcal{L}_{rel}}\frac{\gamma_{\ell}}{\pi_{\ell}},\beta+\max_{\ell\in\mathcal{L}_{rel}}\frac{\gamma_{\ell}}{\pi_{\ell}}\right],
\]
it follows that the \textit{FAS$_{excl}$} is an identified set for
$\beta$ if $0\in\left[\min_{\ell\in\mathcal{L}_{rel}}\frac{\gamma_{\ell}}{\pi_{\ell}},\max_{\ell\in\mathcal{L}_{rel}}\frac{\gamma_{\ell}}{\pi_{\ell}}\right]$.
Therefore, if invalid instruments violate the exclusion assumption
only, then $FAS_{excl}$ is an identified set for $\beta$ if there
is a valid instrument that is relevant, as for that instrument $\frac{\gamma_{\ell}}{\pi_{\ell}}=0$,
or in other words, the invalid instruments have been correctly included
in the model for the valid instrument to point-identify $\beta$.

The sample analog $\hat{FAS}_{excl}$, together with the first-stage hard-thresholding rule for the set of relevant instruments, is given in Appendix~\ref{app:EstiFSHT}.

\section{The \textit{FAS} under Violations of Exogeneity}

\label{sec:FASExo}

MP chose to consider the \textit{FAS} for violations of the exclusion
assumption only. We now consider invalid instruments that act as colliders, such as $Z_3$ in Figure \ref{fig:dag-violations}, whose violation is captured by Definition 2. We employ the MP method for violations of the exogeneity assumption only, and show that it leads to a different \textit{FAS}.

\begin{assumption} \label{Ass:Excl} An invalid instrument $Z_{\ell}$
violates the exogeneity assumption, $\alpha_{\ell}\neq0$. All instruments
satisfy the exclusion assumption, $\gamma_{j}=0$, for $j=1,\ldots,k_{z}$.
\end{assumption}

\noindent As $\boldsymbol{\gamma}=\boldsymbol{0}$, the exogeneity
assumption is no longer conditional on the inclusion of instruments
as controls.

Like the partial exclusion Assumption \ref{Ass:partexcl} of MP, we
make the following partial exogeneity assumption.

\begin{assumption} \label{Ass:PartExo} Partial exogeneity: There
are constants $\eta_{\ell}\geq0$ such that $\left|\alpha_{\ell}\right|\leq\eta_{\ell}$
for $\ell=1,\ldots,k_{z}$.

\end{assumption}

\noindent Under Assumption \ref{Ass:Excl} we have 
\begin{equation}
\boldsymbol{\alpha}=\text{cov}\left(\boldsymbol{Z},Y\right)-\text{cov}\left(\boldsymbol{Z},X\right)\beta.\label{eq:agam0}
\end{equation}
Together with Assumption \ref{Ass:PartExo} we obtain the identified
set for $\beta$ analogous to Theorem 1 in MP for the case where instruments
can violate the exclusion restriction only. We then get that 
\[
\mathcal{B}_{\alpha}\left(\boldsymbol{\eta}\right)=\left\{ b\in\mathbb{R}:-\boldsymbol{\eta}\leq\left(\text{cov}\left(\boldsymbol{Z},Y\right)-\text{cov}\left(\boldsymbol{Z},X\right)b\right)\leq\boldsymbol{\eta}\right\} 
\]
is the identified set for $\beta$, where the inequalities are componentwise.
Consider $\boldsymbol{\eta}\neq\boldsymbol{0}.$ Then for \textbf{$\tilde{\boldsymbol{\eta}}<\boldsymbol{\eta}$,
}the model is falsified if and only if $\mathcal{B}_{\alpha}\left(\tilde{\boldsymbol{\eta}}\right)$
is empty.

Define the $k_{z}$-vectors $\boldsymbol{\pi}^{*}$, $\boldsymbol{\psi}^{*}$,
$\boldsymbol{\alpha}^{*}$ and $\boldsymbol{\eta}^{*}$ with $\ell$-th
elements given by 
\begin{align}
\pi_{\ell}^{*} & \coloneqq\frac{\text{cov}\left(Z_{\ell},X\right)}{\text{var}\left(Z_{\ell}\right)};\,\,\psi_{\ell}^{*}\coloneqq\frac{\text{cov}\left(Z_{\ell},Y\right)}{\text{var}\left(Z_{\ell}\right)};\label{eq:pispsis}\\
\alpha_{\ell}^{*} & \coloneqq\frac{\alpha_{\ell}}{\text{var}\left(Z_{\ell}\right)};\,\,\eta_{\ell}^{*}\coloneqq\frac{\eta_{\ell}}{\text{var}\left(Z_{\ell}\right)},\nonumber 
\end{align}
for $\ell=1,\ldots,k_{z}$. Then it follows that 
\[
\mathcal{B}_{\alpha}\left(\boldsymbol{\eta}\right)=\mathcal{B}_{\alpha^{*}}\left(\boldsymbol{\eta}^{*}\right)=\left\{ b\in\mathbb{R}:-\boldsymbol{\eta}^{*}\leq\left(\boldsymbol{\psi}^{*}-\boldsymbol{\pi}^{*}b\right)\leq\boldsymbol{\eta}^{*}\right\} .
\]
Let $\mathcal{L}_{rel}^{*}$ denote the set of relevant instruments
\begin{equation}
\mathcal{L}_{rel}^{*}=\left\{ \ell\in\left\{ 1,\ldots,k_{z}\right\} :\pi_{\ell}^{*}\neq0\right\} .\label{eq:relsetexo}
\end{equation}
Under Assumptions \ref{Ass:Suff}, \ref{Ass:Excl} and \ref{Ass:PartExo},
and following analogous arguments as in MP, the falsification frontier
is then given as the set 
\[
FF_{exo}=\left\{ \tilde{\boldsymbol{\eta}}^{*}\left(b\right)\in\mathbb{R}_{\geq0}^{k_{z}}:\tilde{\eta}_{\ell}^{*}\left(b\right)=\left|\psi_{\ell}^{*}-b\pi_{\ell}^{*}\right|,\,\ell=1,\ldots,k_{z},\,b\in\left[\min_{\ell\in\mathcal{L}_{rel}^{*}}\frac{\psi_{\ell}^{*}}{\pi_{\ell}^{*}},\max_{\ell\in\mathcal{L}_{rel}^{*}}\frac{\psi_{\ell}^{*}}{\pi_{\ell}^{*}}\right]\right\} .
\]
The resulting falsification adaptive set is then 
\begin{align}
FAS_{exo} & =\left[\min_{\ell\in\mathcal{L}_{rel}^{*}}\frac{\psi_{\ell}^{*}}{\pi_{\ell}^{*}},\max_{\ell\in\mathcal{L}_{rel}^{*}}\frac{\psi_{\ell}^{*}}{\pi_{\ell}^{*}}\right]=\left[\beta+\min_{\ell\in\mathcal{L}_{rel}^{*}}\frac{\alpha_{\ell}^{*}}{\pi_{\ell}^{*}},\beta+\max_{\ell\in\mathcal{L}_{rel}^{*}}\frac{\alpha_{\ell}^{*}}{\pi_{\ell}^{*}}\right].\label{eq:FASexo}
\end{align}

For $\ell\in\mathcal{L}_{rel}^{*}$, the ratio 
\[
\beta_{\ell}^{*}:=\frac{\psi_{\ell}^{*}}{\pi_{\ell}^{*}}=\frac{\text{cov}\left(Z_{\ell},Y\right)}{\text{cov}\left(Z_{\ell},X\right)}
\]
is the IV estimand for $\beta$ in the specification $Y=X\beta+E$,
using $Z_{\ell}$ as the just-identifying instrument for $X$, and
treating the instruments $\boldsymbol{Z}_{\left\{ -\ell\right\} }$
as invalid and excluding them from the analysis. \textit{FAS$_{exo}$}
contains $\beta$ if $0\in\left[\min_{\ell\in\mathcal{L}_{rel}^{*}}\frac{\alpha_{\ell}^{*}}{\pi_{\ell}^{*}},\max_{\ell\in\mathcal{L}_{rel}^{*}}\frac{\alpha_{\ell}^{*}}{\pi_{\ell}^{*}}\right]$.
Therefore, if invalid instruments violate the exogeneity assumption
only, then $FAS_{exo}$ is an identified set for $\beta$ if there
is a valid instrument that is relevant, as for that instrument $\frac{\alpha_{\ell}^{*}}{\pi_{\ell}^{*}}=0$.
Again, for this case invalid instruments are correctly excluded from
the instrument set for the valid instrument to point-identify $\beta.$

If one believes that invalid instruments violate the exclusion assumption only, acting as confounders, \textit{FAS}$_{excl}$ is the appropriate set reflecting the model uncertainty that arises from a falsified baseline model. This would for example be the case when instruments are randomly assigned, but some have a direct effect on the outcome. If instead the researcher believes the exclusion assumption to hold on economic grounds, but questions the exogeneity assumption of some of the instruments due to no or improper randomization, \textit{FAS}$_{exo}$, which treats invalid IVs as colliders, would be the appropriate set. If there is a valid and relevant IV, the falsification adaptive set is guaranteed to be an identified set for $\beta$ if the invalid instruments are treated appropriately, and hence the choice of maintained assumption is consequential.

The two falsification frontiers and adaptive sets derived above are
simply based on just-identified estimands, and below we will further
exploit this to consider different combinations of violations of the
exclusion and conditional exogeneity assumptions. It is at this point
important to stress that these falsification frontiers are frontiers
no matter what the actual violations of the exclusion or conditional
exogeneity assumptions are. Also, exclusion violations imply exogeneity
violations and vice versa, as when $\text{var}\left(\boldsymbol{Z}\right)^{-1}\left(\text{cov}\left(\boldsymbol{Z},Y-X\beta\right)\right)=\boldsymbol{\gamma}$
it follows that $\boldsymbol{\alpha}\left(\gamma\right)=\text{cov}\left(\boldsymbol{Z},Y-X\beta\right)=\text{var}\left(\boldsymbol{Z}\right)\boldsymbol{\gamma}$
and when $\text{cov}\left(\boldsymbol{Z},Y-X\beta\right)=\boldsymbol{\alpha}$
it follows that $\boldsymbol{\gamma}\left(\alpha\right)=$$\text{var}\left(\boldsymbol{Z}\right)^{-1}\boldsymbol{\alpha}$.
MP used this latter observation to argue that their \textit{FAS}$_{excl}$
also applies to the violations of the exogeneity assumption only.
Further, in the general situation with $E=\boldsymbol{Z}'\boldsymbol{\gamma}+U$,
$\text{cov}\left(\boldsymbol{Z},U\right)=\boldsymbol{\alpha}$ we
have $\boldsymbol{\gamma}\left(\alpha\right)=\boldsymbol{\gamma}+\text{var}\left(\boldsymbol{Z}\right)^{-1}\boldsymbol{\alpha}$
and $\boldsymbol{\alpha}\left(\gamma\right)=\boldsymbol{\alpha}+\text{var}\left(\boldsymbol{Z}\right)\boldsymbol{\gamma}$.

For now, we have these two falsification frontiers and associated
falsification adaptive sets. MP (p. 1449) characterised the falsification
frontier as the set of smallest relaxations of the baseline model
which are not falsified. But with multiple falsification frontiers,
their exclusion based falsification frontier may not be the set of
smallest relaxations. As a simple illustration, consider the following
numerical example. 
\begin{example}
\label{example:fas-1} We set $k_{z}=2$, $\text{\ensuremath{\beta=}}\frac{1}{3}$
and 
\[
\boldsymbol{\gamma}=\left(\begin{array}{c}
-1\\
1
\end{array}\right);\,\,\,\text{var}\left(\boldsymbol{Z}\right)=\left[\begin{array}{cc}
1 & 0.5\\
0.5 & 1
\end{array}\right];\,\,\,\text{cov}\left(\boldsymbol{Z},X\right)=\left(\begin{array}{c}
1.5\\
1.5
\end{array}\right).
\]
Then 
\[
\text{cov}\left(\boldsymbol{Z},Y\right)=\left(\begin{array}{c}
0\\
1
\end{array}\right);\,\,\,\boldsymbol{\pi}=\left(\begin{array}{c}
1\\
1
\end{array}\right);\,\,\,\boldsymbol{\psi}=\left(\begin{array}{c}
-\frac{2}{3}\\
\frac{4}{3}
\end{array}\right);\,\,\boldsymbol{\pi}^{*}=\left(\begin{array}{c}
1.5\\
1.5
\end{array}\right);\,\,\,\boldsymbol{\psi}^{*}=\left(\begin{array}{c}
0\\
1
\end{array}\right).
\]

For \textit{FAS}$_{exo}$, we have 
\[
\boldsymbol{\alpha}\left(\gamma\right)\coloneqq\text{cov}\left(\boldsymbol{Z},Y-X\beta\right)=\text{var}\left(\boldsymbol{Z}\right)\boldsymbol{\gamma}=\left(\begin{array}{c}
	-\frac{1}{2}\\
	\frac{1}{2}
\end{array}\right).
\]
We thus have for $\eta_{1}\geq\frac{1}{2}$ and $\eta_{2}\geq\frac{1}{2}$
\[
\left(\begin{array}{c}
	-\eta_{1}\\
	-\eta_{2}
\end{array}\right)\leq\left(\begin{array}{c}
	\alpha\left(\gamma\right)_{1}\\
	\alpha\left(\gamma\right)_{2}
\end{array}\right)\leq\left(\begin{array}{c}
	\eta_{1}\\
	\eta_{2}
\end{array}\right)
\]
and the identified set is given by 
\[
\mathcal{B}_{\alpha}\left(\boldsymbol{\eta}\right)=\left[\max\left(-\frac{2}{3}\eta_{1},\frac{2}{3}-\frac{2}{3}\eta_{2}\right),\min\left(\frac{2}{3}\eta_{1},\frac{2}{3}+\frac{2}{3}\eta_{2}\right)\right].
\]
It follows that $\mathcal{B}_{\alpha}\left(\tilde{\boldsymbol{\eta}}\right)$
is empty when $\tilde{\eta}_{1}+\tilde{\eta}_{2}<1$. The falsification
frontier is obtained for values $\tilde{\eta}_{1}+\tilde{\eta}_{2}=1$,
and 
\[
FAS_{exo}=\left[0,\frac{2}{3}\right]=\left[\frac{\psi_{1}^{*}}{\pi_{1}^{*}},\frac{\psi_{2}^{*}}{\pi_{2}^{*}}\right].
\]

The analogous calculation for the falsification
frontier based on the exclusion restriction is obtained for values $\tilde{\delta}_{1}+\tilde{\delta}_{2}=2$,
as
\[
FAS_{excl}=\left[-\frac{2}{3},\frac{4}{3}\right]=\left[\frac{\psi_{1}}{\pi_{1}},\frac{\psi_{2}}{\pi_{2}}\right].
\]
\end{example}

In this example both falsification adaptive sets
include $\beta=\frac{1}{3}$, as the two elements of
both $\boldsymbol{\pi}$ and $\boldsymbol{\pi}^{*}$ have equal signs,
whereas those of $\boldsymbol{\gamma}$ and $\boldsymbol{\alpha}\left(\gamma\right)$
have opposite signs. Of the two, \textit{FAS}$_{exo}$ is the much
smaller set, indicating
a smaller relaxation of the moment conditions.

However, when we change the values of $\boldsymbol{\gamma}$ to $\boldsymbol{\gamma}=\left(0,1\right)'$
whilst keeping the values of $\beta$, $\text{var}\left(\boldsymbol{Z}\right)$
and $\text{cov}\left(\boldsymbol{Z},X\right)$ the same, we obtain
$FAS_{excl}=\left[\frac{1}{3},\frac{4}{3}\right]$ and $FAS_{exo}=\left[\frac{2}{3},1\right]$.
Now \textit{FAS}$_{excl}$ contains $\beta$ whereas \textit{FAS}$_{exo}$
does not, and \textit{FAS}$_{exo}$ is again the smaller set. $Z_{1}$
is here a valid and relevant instrument, but due to the correlation
of $Z_{1}$ and $Z_{2}$, only identifies $\beta$ when $Z_{2}$ is
included in the model as a control. This is one of the estimands of
\textit{FAS}$_{excl}$, as detailed in equation (\ref{eq:jidcon}),
but not of \textit{FAS}$_{exo}$. Another way to look at it is that
here $\boldsymbol{\alpha}\left(\gamma\right)=\text{var}\left(\boldsymbol{Z}\right)\boldsymbol{\gamma}=\left(0.5,1\right)'$,
and so although there is a valid instrument, both elements of $\boldsymbol{\alpha}\left(\gamma\right)$
are non-zero due to the correlation of the instruments. Both elements
are here positive, so $\beta\notin FAS_{exo}$. \textit{FAS}$_{excl}$
is guaranteed to contain $\beta$ under Assumption \ref{Ass:Exo}
that invalid instruments violate the exclusion assumption only if
there is at least one valid and relevant instrument $Z_{\ell}$ with
$\gamma_{\ell}=0$ and $\pi_{\ell}\neq0$. Likewise, $FAS_{exo}$
is guaranteed to contain $\beta$ under Assumption \ref{Ass:Excl}
that invalid instruments violate the exogeneity assumption only if
there is at least one valid and relevant instrument $Z_{\ell}$ with
$\alpha_{\ell}=0$ and $\pi_{\ell}^{*}\neq0$.

This confirms the consequential role of the maintained assumption: the choice of \textit{FAS} must match the type of violation, and selecting the wrong set can exclude $\beta$ even when a valid instrument is present. The sample analog $\hat{FAS}_{exo}$ is given in Appendix~\ref{app:EstiFASexo}.

\section{Alternative Derivation}

\label{sec:Alternative-Derivation}

To derive the \textit{FF} and \textit{FAS} in a different way and
without starting from the identified sets from the partial relaxations,
consider the population 2SLS criterion 
\[
S\left(b\right)=\text{cov}\left(\boldsymbol{Z},Y-Xb\right)'\left(\text{var}\left(\boldsymbol{Z}\right)\right)^{-1}\text{cov}\left(\boldsymbol{Z},Y-Xb\right).
\]
If $\text{cov}\left(\boldsymbol{Z},Y-X\beta\right)=\boldsymbol{0}$,
then clearly $\arg\min_{b}S\left(b\right)=\beta$ and $S\left(\beta\right)=0$,
and the model is not falsified. If $\text{cov}\left(\boldsymbol{Z},Y-X\beta\right)=\boldsymbol{\alpha}$
and if we know $\boldsymbol{\alpha}$, then we get the same results
for 
\[
S\left(b;\boldsymbol{\alpha}\right)=\left(\text{cov}\left(\boldsymbol{Z},Y-Xb\right)-\boldsymbol{\alpha}\right)'\left(\text{var}\left(\boldsymbol{Z}\right)\right)^{-1}\left(\text{cov}\left(\boldsymbol{Z},Y-Xb\right)-\boldsymbol{\alpha}\right).
\]
But of course, we don't know $\boldsymbol{\alpha}$. The falsification
frontier is obtained as follows. Let, for $\ell\in\mathcal{L}_{rel}^{*}$,
\[
\beta_{\ell}^{*}=\text{cov}\left(Z_{\ell},Y\right)/\text{cov}\left(Z_{\ell},X\right)
\]
and 
\[
\boldsymbol{a}\left(\beta_{\ell}^{*}\right)=\text{cov}\left(\boldsymbol{Z},Y-X\beta_{\ell}^{*}\right),
\]
so $\boldsymbol{a}\left(\beta_{\ell}^{*}\right)_{\ell}=0$. It follows
that for 
\[
S\left(b;\boldsymbol{a}\left(\beta_{\ell}^{*}\right)\right)=\left(\text{cov}\left(\boldsymbol{Z},Y-Xb\right)-\boldsymbol{a}\left(\beta_{\ell}^{*}\right)\right)'\left(\text{var}\left(\boldsymbol{Z}\right)\right)^{-1}\left(\text{cov}\left(\boldsymbol{Z},Y-Xb\right)-\boldsymbol{a}\left(\beta_{\ell}^{*}\right)\right),
\]
we get 
\[
\arg\min_{b}S\left(b;\boldsymbol{a}\left(\beta_{\ell}^{*}\right)\right)=\beta_{\ell}^{*};\,\,S\left(\beta_{\ell}^{*};\boldsymbol{a}\left(\beta_{\ell}^{*}\right)\right)=0,
\]
for $\ell=1,\ldots,k_{z}$. So for each just-identifying IV estimand
$\beta_{\ell}^{*}$ we adjust the other moment conditions by the amount
needed to get $\text{cov}\left(\boldsymbol{Z},Y-X\beta_{\ell}^{*}\right)-\boldsymbol{a}\left(\beta_{\ell}^{*}\right)=\boldsymbol{0}$,
with $\boldsymbol{a}\left(\beta_{\ell}^{*}\right)_{\ell}=0$.

It also follows that for $\beta^{*}\in\left[\min_{\ell\in\mathcal{L}_{rel}^{*}}\beta_{\ell}^{*},\max_{\ell\in\mathcal{L}_{rel}^{*}}\beta_{\ell}^{*}\right]$,
we have $\arg\min_{b}S\left(b;\boldsymbol{a}\left(\beta^{*}\right)\right)=\beta^{*}$
and $S\left(\beta^{*};\boldsymbol{a}\left(\beta^{*}\right)\right)=0$,
as this is true for any $\beta^{*}.$

This is a falsification frontier, as when moving $\beta^{*}$ from
$\min_{\ell\in\mathcal{L}_{rel}^{*}}\beta_{\ell}^{*}$ to $\max_{\ell\in\mathcal{L}_{rel}^{*}}\beta_{\ell}^{*}$,
there is at least one element in $\left|\boldsymbol{a}\left(\beta^{*}\right)\right|$
that decreases in value and at least one that increases in value,
and any $\left|\boldsymbol{a}\right|<\left|\boldsymbol{a}\left(\beta^{*}\right)\right|$
results in a falsification of the model.

The resulting $FAS$ is $FAS_{exo}=\left[\min_{\ell\in\mathcal{L}_{rel}^{*}}\beta_{\ell}^{*},\max_{\ell\in\mathcal{L}_{rel}^{*}}\beta_{\ell}^{*}\right]$,
which is the set 
\[
\mathcal{B}\left(\tilde{\boldsymbol{\eta}}_{F}\right)=\left\{ b\in\mathbb{R}:-\tilde{\boldsymbol{\eta}}_{F}\leq\left(\text{cov}\left(\boldsymbol{Z},Y\right)-\text{cov}\left(\boldsymbol{Z},X\right)b\right)\leq\tilde{\boldsymbol{\eta}}_{F}\right\} ,
\]
where 
\[
\tilde{\eta}_{F,j}=\max_{\ell\in\mathcal{L}_{rel}^{*}}\left|\boldsymbol{a}\left(\beta_{\ell}^{*}\right)_{j}\right|=\max_{\ell\in\mathcal{L}_{rel}^{*}}\left|\text{cov}\left(Z_{j},X\right)\left(\beta_{j}^{*}-\beta_{\ell}^{*}\right)\right|.
\]
We added the subscript $F$ to make clear that these relaxations come
from the falsification frontier. Again, it is clear that it could
be the case that $\beta\notin\mathcal{B}\left(\tilde{\boldsymbol{\eta}}_{F}\right)$.

The argument is symmetric for $FAS_{excl}$. We can write the population 2SLS criterion equivalently as
\[
\begin{aligned}S\left(b\right) & =\text{cov}\left(\boldsymbol{Z},Y-Xb\right)'\left(\text{var}\left(\boldsymbol{Z}\right)\right)^{-1}\text{var}\left(\boldsymbol{Z}\right)\left(\text{var}\left(\boldsymbol{Z}\right)\right)^{-1}\text{cov}\left(\boldsymbol{Z},Y-Xb\right)\\
 & =\left(\boldsymbol{\psi}-\boldsymbol{\pi}b\right)'\text{var}\left(\boldsymbol{Z}\right)\left(\boldsymbol{\psi}-\boldsymbol{\pi}b\right),\end{aligned}
\]
and if $\boldsymbol{\psi}-\boldsymbol{\pi}\beta=\boldsymbol{\gamma}$, then for
\[
S\left(b;\boldsymbol{\gamma}\right)=\left(\boldsymbol{\psi}-\boldsymbol{\pi}b-\boldsymbol{\gamma}\right)'\text{var}\left(\boldsymbol{Z}\right)\left(\boldsymbol{\psi}-\boldsymbol{\pi}b-\boldsymbol{\gamma}\right),
\]
we have $\arg\min_{b}S\left(b;\boldsymbol{\gamma}\right)=\beta$ and $S\left(\beta;\boldsymbol{\gamma}\right)=0$. Applying the same arguments as above with $\boldsymbol{c}\left(\beta_{\ell}\right)=\boldsymbol{\psi}-\boldsymbol{\pi}\beta_{\ell}$ for $\ell\in\mathcal{L}_{rel}$ yields $FAS_{excl}=\left[\min_{\ell\in\mathcal{L}_{rel}}\beta_{\ell},\max_{\ell\in\mathcal{L}_{rel}}\beta_{\ell}\right]$, and again $\beta$ may not be contained in this set.

From these derivations it is clear that the falsification frontiers
are bounded relaxations of the moment conditions that lead to nonfalsification,
independent of the actual types of violations of the exclusion or
conditional exogeneity assumptions.

\section{Combining Exclusion and Exogeneity Violations}

\label{sec:ExoExcl}

Up to this point, we have assumed that invalid instruments either all violate the exclusion assumption (Section \ref{sec:FASExcl}) or all violate the exogeneity assumption (Section \ref{sec:FASExo}). In practice, however, a researcher may not know which assumption each invalid instrument violates, and there is no reason to expect a clear-cut case where all invalid instruments are of the same type. Some may act as confounders, others as colliders. In the roads and trade example of DMT, the 1947 highways instrument may plausibly act as a confounder while the historical exploration routes instrument may act as a collider. These different patterns of violations matter greatly for the \textit{FAS}: as we show below, each pattern gives rise to a different pattern-specific \textit{FAS}. Crucially, whether the true $\beta$ is contained in the \textit{FAS} hinges on the choice of pattern.

If the researcher wishes to remain agnostic about which invalid instruments are confounders and which are colliders, a generalized \textit{FAS} that accounts for all possible patterns is needed. This may well be the most plausible setting in the DMT application, where the researcher cannot be certain which instrument violates which assumption.

\subsection{Patterns of Violations}

From the results above, it is clear that one gets a different \textit{FF
}and \textit{FAS} for any nonsingular linear transformation of the
instruments, $\boldsymbol{Z}^{*}=\boldsymbol{A}\boldsymbol{Z}$, with
such transformations leading to identical estimation and falsification
test results. For the violation of the exclusion assumption only,
this transformation is given by $\boldsymbol{Z}^{*}=\left(\text{var}\left(\boldsymbol{Z}\right)\right)^{-1}\boldsymbol{Z}$.
The $\ell^{th}$ element for this $\boldsymbol{Z}^{*}$ can be written
as 
\[
Z_{\ell}^{*}=\left(\text{var}\left(\tilde{Z}_{\ell}\right)\right)^{-1}\tilde{Z}_{\ell},
\]
where 
\[
\tilde{Z}_{\ell}=Z_{\ell}-\boldsymbol{Z}_{\left\{ -\ell\right\} }^{\prime}\left(\text{var}\left(\boldsymbol{Z}_{\left\{ -\ell\right\} }\right)\right)^{-1}\text{cov}\left(\boldsymbol{Z}_{\left\{ -\ell\right\} },Z_{\ell}\right)
\]
is the population residual after linearly partialling out $\boldsymbol{Z}_{\left\{ -\ell\right\} }$
from $Z_{\ell}$. It is the population just-identifying instrument
after including $\boldsymbol{Z}_{\left\{ -\ell\right\} }$ as explanatory
variables, as in model specification (\ref{eq:jidcon}), and we write
such $\tilde{Z}_{\ell}$ equivalently as $Z_{\ell|\left\{ -\ell\right\} }$.
It follows that $\text{cov}\left(Z_{\ell}^{*}\left(Y-X\beta\right)\right)=\gamma_{\ell}$.

Likewise, for the violation of the exogeneity assumption only, we
can specify $Z_{\ell}^{*}=\left(\text{var}\left(Z_{\ell}\right)\right)^{-1}Z_{\ell}$,
resulting in $\text{cov}\left(Z_{\ell}^{*}\left(Y-X\beta\right)\right)=\alpha_{\ell}^{*}$
as in specification (\ref{eq:pispsis}).

We can from here consider nonsingular transformations $\boldsymbol{Z}^{*}=\boldsymbol{A}\boldsymbol{Z}$
that span possible patterns of violations of the exclusion and conditional
exogeneity assumptions. We consider cases where there can be a mixture
of invalid instruments, with some invalid instruments violating the
exclusion assumption and some the conditional exogeneity assumption:

\begin{assumption} \label{Ass:exoexcl} An invalid instrument $Z_{\ell}$
violates either the exclusion assumption, $\gamma_{\ell}\neq0$, or
the conditional exogeneity assumption, $\alpha_{\ell}\neq0$, but
not both, $\gamma_{\ell}\alpha_{\ell}=0$. \end{assumption}

\noindent Assumption \ref{Ass:exoexcl} is a generalization that includes
the previous two cases of violating the exclusion or exogeneity assumption
only (Assumptions \ref{Ass:Exo} and \ref{Ass:Excl}), and which enables
us to define a joint relaxation of the exclusion and conditional exogeneity
assumptions.

We consider natural relaxations of the exclusion and conditional exogeneity
assumptions for each possible pattern that $\boldsymbol{\gamma}+\boldsymbol{\alpha}$
can take with $\gamma_{\ell}\alpha_{\ell}=0$ for $\ell=1,\ldots,k_{z}$.
There are $S=2^{k_{z}}$ such patterns. Denote these patterns by $\boldsymbol{\gamma}_{s}+\boldsymbol{\alpha}_{s}$,
$s=1,\ldots,S$. Let $\mathcal{K}=\left\{ 1,\ldots,k_{z}\right\} $
and let $\mathcal{C}$ denote the collection of all $S$ possible
subsets of $\mathcal{K}$, so $\mathcal{C}=\left\{ \emptyset,1,\ldots,\left\{ 1,2\right\} ,\ldots,\mathcal{K}\right\} $.
Denote the subsets of $\mathcal{C}$ by $\mathcal{C}_{s}$, $s=1,\ldots,S$,
and let $\mathcal{A}_{s}=\mathcal{K}\setminus\mathcal{C}_{s}$. Then
we consider all possible patterns as follows. For $s=1,\ldots,S$,
for an invalid instrument $Z_{\ell}$, $\ell\in\mathcal{K}$, if $\ell\in\mathcal{C}_{s}$
it violates the exclusion restriction, $\gamma_{\ell}\neq0$, and
if $\ell\in\mathcal{A}_{s}$ it violates the conditional exogeneity
assumption, $\alpha_{\ell}\neq0$. To illustrate for $k_{z}=3$, with
$\boldsymbol{\gamma}_{s}+\boldsymbol{\alpha}_{s}=\left(\alpha_{1},\gamma_{2},\alpha_{3}\right)'$,
the associated sets are $\mathcal{C}_{s}=\left\{ 2\right\} $ and
$\mathcal{A}_{s}=\left\{ 1,3\right\} $. Further, we obtain a violation
of the exclusion assumption only, $\boldsymbol{\gamma}_{s}+\boldsymbol{\alpha}_{s}=\boldsymbol{\gamma}$,
for $\mathcal{C}_{s}=\mathcal{K}$, $\mathcal{A}_{s}=\emptyset$,
and a violation of the exogeneity assumption only, $\boldsymbol{\gamma}_{s}+\boldsymbol{\alpha}_{s}=\boldsymbol{\alpha}$,
for $\mathcal{C}_{s}=\mathcal{\emptyset}$, $\mathcal{A}_{s}=\mathcal{K}$.
Note that we consider all possible patterns where all instruments
can be invalid, but for a valid instrument $Z_{\ell}$ we have that
$\gamma_{\ell}=\alpha_{\ell}=0$ as per Definition \ref{Def:ValIInst}.

\subsection{The Generalized FAS}

We can now make the following natural joint relaxation of the exclusion
and conditional exogeneity assumptions.

\begin{assumption} \label{Ass:PartExclExo}

Partial joint exclusion and conditional exogeneity: For each pattern
$\boldsymbol{\gamma}_{s}+\boldsymbol{\alpha}_{s}$, $s=1,\ldots,S$,
with $S=2^{k_{z}}$ and $\gamma_{s,\ell}\alpha_{s,\ell}=0$ for $\ell=1,\ldots,k_{z}$,
there are constants $\omega_{s,\ell}\geq0$ such that $\left|\gamma_{s,\ell}+\alpha_{s,\ell}\right|\leq\omega_{s,\ell}$
for $\ell=1,\ldots,k_{z}$.

\end{assumption}

\noindent For each $s\in\left\{ 1,\ldots,S\right\} $ we obtain the
identified set, falsification frontier and falsification adaptive
set. For pattern $\left(\mathcal{C}_{s},\mathcal{A}_{s}\right)$ we
have the model specification 
\[
Y=X\beta+\boldsymbol{Z}_{\mathcal{C}_{s}}^{\prime}\boldsymbol{\gamma}_{\mathcal{C}_{s}}+\tilde{U};\,\,\,\text{cov}\left(\boldsymbol{Z}_{\mathcal{A}_{s}},\tilde{U}\right)=\boldsymbol{\alpha}_{\mathcal{A}_{s}}.
\]
As a unifying framework and following the discussion above, we consider
linearly transformed instruments as follows. For $\ell\in\mathcal{C}_{s}$,
let $\mathcal{C}_{s,-\ell}=\mathcal{C}_{s}\setminus\left\{ \ell\right\} $.
Then, for $\ell\in\mathcal{C}_{s}$, we linearly partial out $\boldsymbol{Z}_{\mathcal{C}_{s,-\ell}}$
from $Z_{\ell}$, 
\[
Z_{\ell|\mathcal{C}_{s,\left\{ -\ell\right\} }}=Z_{\ell}-\boldsymbol{Z}_{\mathcal{C}_{s,-\ell}}^{\prime}\left(\text{var}\left(\boldsymbol{Z}_{\mathcal{C}_{s,-\ell}}\right)\right)^{-1}\text{cov}\left(\boldsymbol{Z}_{\mathcal{C}_{s,-\ell}},Z_{\ell}\right)
\]
It then follows that 
\[
\frac{\text{cov}\left(Z_{\ell|\mathcal{C}_{s,-\ell}},Y-X\beta\right)}{\text{var}\left(Z_{\ell|\mathcal{C}_{s,-\ell}}\right)}=\gamma_{s,\ell},\,\,\,\forall\ell\in\mathcal{C}_{s}.
\]
Likewise, for $\ell\in\mathcal{A}_{s}$, we partial out $\boldsymbol{Z}_{\mathcal{C}_{s}}$
from $Z_{\ell}$, 
\[
Z_{\ell|\mathcal{C}_{s}}=Z_{\ell}-\boldsymbol{Z}_{\mathcal{C}_{s}}^{\prime}\left(\text{var}\left(\boldsymbol{Z}_{\mathcal{C}_{s}}\right)\right)^{-1}\text{cov}\left(\boldsymbol{Z}_{\mathcal{C}_{s}},Z_{\ell}\right),
\]
resulting in 
\[
\frac{\text{cov}\left(Z_{\ell|\mathcal{C}_{s}},Y-X\beta\right)}{\text{var}\left(Z_{\ell|\mathcal{C}_{s}}\right)}=\frac{\alpha_{s,\ell}}{\text{var}\left(Z_{\ell|\mathcal{C}_{s}}\right)}\equiv\tilde{\alpha}_{s,\ell},\,\,\,\forall\ell\in\mathcal{A}_{s}.
\]

\noindent Then define $\tilde{\boldsymbol{Z}}_{s}$ as the $k_{z}$-vector
with $\ell$-th element either $Z_{\ell|\mathcal{C}_{s,-\ell}}$ if
$\ell\in\mathcal{\mathcal{C}}_{s}$, or $Z_{\ell|\mathcal{C}_{s}}$
if $\ell\in\mathcal{A}_{s}$. For our example with $\boldsymbol{\gamma}_{s}+\boldsymbol{\alpha}_{s}=\left(\alpha_{1},\gamma_{2},\alpha_{3}\right)'$,
we have $\tilde{\boldsymbol{Z}}_{s}=\left(Z_{1|2},Z_{2},Z_{3|2}\right)'$.
Let 
\[
\boldsymbol{Z}^{*}=\left(\text{diag}\left(\text{var}\left(\tilde{\boldsymbol{Z}}_{s}\right)\right)\right)^{-1}\tilde{\boldsymbol{Z}}_{s},
\]
where for a general square matrix $\boldsymbol{Q}$, $\text{diag}\left(\boldsymbol{Q}\right)$
is a diagonal matrix containing the diagonal elements of $\boldsymbol{Q}$.
It then follows that 
\[
\text{cov}\left(\boldsymbol{Z}_{s}^{*},Y-X\beta\right)=\boldsymbol{\gamma}_{s}+\tilde{\boldsymbol{\alpha}}_{s},
\]
where $\tilde{\boldsymbol{\alpha}}_{s}$ is the $k_{z}$-vector with
$\ell$-th element either $\tilde{\alpha}_{s,\ell}$ if $\ell\in\mathcal{A}_{s}$,
or $0$ if $\ell\in\mathcal{C}_{s}$. Let \textbf{$\tilde{\boldsymbol{\omega}}$}
denote the $k_{z}$-vector with $\ell$-th element either $\text{var}\left(\tilde{Z}_{s,\ell}\right)^{-1}\omega_{s,\ell}$
if $\ell\in\mathcal{A}_{s}$, or $\omega_{s,\ell}$ if $\ell\in\mathcal{C}_{s}$,
then it follows from Assumption \ref{Ass:PartExclExo} that $-\tilde{\boldsymbol{\omega}}_{s}\leq\boldsymbol{\gamma}_{s}+\tilde{\boldsymbol{\alpha}}_{s}\leq\tilde{\boldsymbol{\omega}}_{s}$.
For each $s\in\left\{ 1,\ldots,S\right\} $ we then get the identified
set as detailed in the following proposition. 
\begin{prop}
\label{Prop:IdSetExoExcl} Suppose Assumptions \ref{Ass:Suff}, \ref{Ass:exoexcl}
and \ref{Ass:PartExclExo} hold, then, for each $s\in\left\{ 1,\ldots,S\right\} $,
\[
\mathcal{B}_{s}\left(\tilde{\boldsymbol{\omega}}_{s}\right)=\left\{ b\in\mathbb{R}:-\tilde{\boldsymbol{\omega}}_{s}\leq\text{cov}\left(\boldsymbol{Z}_{s}^{*},Y-Xb\right)\leq\tilde{\boldsymbol{\omega}}_{s}\right\} ,
\]
is the sharp identified set for $\beta$, where the inequalities are
componentwise, and where the definitions of $\boldsymbol{Z}_{s}^{*}$
and $\tilde{\boldsymbol{\omega}}_{s}$ are given in the preceding
text. For each $s$, the model is falsified if and only if this set
is empty. 
\end{prop}

\noindent The proof follows the same arguments as those of Theorem
1 in MP, see \citet{Apfel2022}.

For $s=1\ldots,S$, define the $k_{z}$-vectors $\tilde{\boldsymbol{\pi}}_{s}$
and $\tilde{\boldsymbol{\psi}}_{s}$, with $\ell$-th elements given
by 
\begin{align}
\tilde{\pi}_{s,\ell} & \coloneqq\frac{\text{cov}\left(\tilde{Z}_{s,\ell},X\right)}{\text{var}\left(\tilde{Z}_{s,\ell}\right)};\,\,\tilde{\psi}_{s,\ell}\coloneqq\frac{\text{cov}\left(\tilde{Z}_{s,\ell},Y\right)}{\text{var}\left(\tilde{Z}_{s,\ell}\right)},\label{eq:pipsisl}
\end{align}
for $\ell=1,\ldots,k_{z}$. Then it follows that 
\[
\mathcal{B}_{s}\left(\tilde{\boldsymbol{\omega}}\right)=\left\{ b\in\mathbb{R}:-\tilde{\boldsymbol{\omega}}_{s}\leq\left(\tilde{\boldsymbol{\psi}}_{s}-\tilde{\boldsymbol{\pi}}_{s}b\right)\leq\tilde{\boldsymbol{\omega}}_{s}\right\} .
\]
Let $\tilde{\mathcal{L}}_{s,rel}$ denote the set of relevant instruments
\begin{equation}
\tilde{\mathcal{L}}_{s,rel}=\left\{ \ell\in\left\{ 1,\ldots,k_{z}\right\} :\tilde{\pi}_{s,\ell}\neq0\right\} .\label{eq:relsetexoexcl}
\end{equation}
Under Assumptions \ref{Ass:Suff}, \ref{Ass:exoexcl} and \ref{Ass:PartExclExo},
the falsification frontier for $s\in\left\{ 1\ldots,S\right\} $,
is then given by 
\[
FF_{s}=\left\{ \tilde{\tilde{\boldsymbol{\omega}}}_{s}\left(b\right)\in\mathbb{R}_{\geq0}^{k_{z}}:\tilde{\tilde\omega}_{s,\ell}\left(b\right)=\left|\tilde{\psi}_{s,\ell}-b\tilde{\pi}_{s,\ell}\right|,\,\ell=1,\ldots,k_{z},\,b\in\left[\min_{\ell\in\tilde{\mathcal{L}}_{s,rel}}\frac{\tilde{\psi}_{s,\ell}}{\tilde{\pi}_{s,\ell}},\max_{\ell\in\tilde{\mathcal{L}}_{s,rel}}\frac{\tilde{\psi}_{s,\ell}}{\tilde{\pi}_{s,\ell}}\right]\right\} ,
\]
and the resulting falsification adaptive set is 
\begin{align}
FAS_{s} & =\left[\min_{\ell\in\tilde{\mathcal{L}}_{s,rel}}\frac{\tilde{\psi}_{s,\ell}}{\tilde{\pi}_{s,\ell}},\max_{\ell\in\tilde{\mathcal{L}}_{s,rel}}\frac{\tilde{\psi}_{s,\ell}}{\tilde{\pi}_{s,\ell}}\right].\label{eq:FASexclexos}
\end{align}

Here, 
\[
\tilde{\beta}_{s,\ell}\coloneqq\frac{\tilde{\psi}_{s,\ell}}{\tilde{\pi}_{s,\ell}}=\frac{\text{cov}\left(\tilde{Z}_{s,\ell},Y\right)}{\text{cov}\left(\tilde{Z}_{s,\ell},X\right)}
\]
is the IV estimand for $\beta$ in the specification $Y=X\beta+U$,
using the transformed instrument $\tilde{Z}_{s,\ell}$ as the just-identifying
instrument for $X$, but these imply different model specifications.
For our $k_{z}=3$ example with $\boldsymbol{\gamma}+\boldsymbol{\alpha}=\left(\alpha_{1},\gamma_{2},\alpha_{3}\right)'$,
we have that $\tilde{\beta}_{1|2}=\frac{\text{cov}\left(Z_{1|2},Y\right)}{\text{cov}\left(Z_{1|2},X\right)}$
is the IV estimand for $\beta$ in the model with $Z_{2}$ included
as a control and $Z_{3}$ excluded from the instrument set. It follows
that for this example $\tilde{\beta}_{1|2}=\beta+\frac{\tilde{\alpha}_{1}}{\pi_{1|2}}$.

For each $s$, the estimands $\tilde{\beta}_{s,\ell}$ are based on
the just-identifying instruments $\tilde{Z}_{s,l}$, for $\ell=1,\ldots,k_{z}$.
Therefore the alternative derivation of the \textit{FF} and \textit{FAS}
as given in Section \ref{sec:Alternative-Derivation} by means of
the population 2SLS criterion function applies directly, as we consider
the nonsingular transformations $\boldsymbol{Z}_{s}^{*}=\boldsymbol{A}_{s}\boldsymbol{Z}$.
These span the set of $S$ possible combinations of violations of
the exclusion and conditional exogeneity assumption under Assumption
\ref{Ass:exoexcl}.

For the $k_{z}=3$ case, if we consider violations of the exclusion
assumption only, $\boldsymbol{\gamma}+\boldsymbol{\alpha}=\boldsymbol{\text{\ensuremath{\gamma}}}=\left(\gamma_{1},\gamma_{2},\gamma_{3}\right)'$,
we have $\tilde{\boldsymbol{Z}}_{s}=\left(Z_{1|23},Z_{2|13},Z_{3|12}\right)'$,
resulting in \textit{FAS}$_{excl}$. For violations of the exogeneity
assumption only, $\boldsymbol{\gamma}+\boldsymbol{\alpha}=\boldsymbol{\text{\ensuremath{\alpha}}}=\left(\alpha_{1},\alpha_{2},\alpha_{3}\right)'$,
we have $\tilde{\boldsymbol{Z}}_{s}=\left(Z_{1},Z_{2},Z_{3}\right)'$,
resulting in \textit{FAS}$_{exo}$. Note that if $\text{var}\left(\boldsymbol{Z}\right)$
is a diagonal matrix then all \textit{FAS$_{s}$}, $s=1,\ldots,S$,
are identical.

We now define the generalized \textit{FAS} as follows.

\begin{defi} \label{Def:GenFAS}

Generalized Falsification Adaptive Set: For $s=1,\ldots,S$, let \textit{FAS}$_{s}$
be as defined in (\ref{eq:FASexclexos}). The generalized falsification
adaptive set is the union of the pattern-specific falsification adaptive
sets, 
\begin{equation}
FAS_{G}=\cup_{s=1}^{S}FAS_{s}.\label{eq:FASrel}
\end{equation}

\end{defi}

\noindent The generalized \textit{FAS} reflects the model uncertainty
that arises from a falsified baseline model, considering all possible
patterns of violations of the exclusion and conditional exogeneity
assumptions that satisfy Assumption \ref{Ass:exoexcl}.

The motivation for introducing the generalized \textit{FAS} is that
when a set of relaxed assumptions does not match the true violation
patterns, the respective $FAS_{s}$ might not include the true $\beta$
even when there is a valid instrument $Z_{\ell}$ with $\gamma_{\ell}=\alpha_{\ell}=0$.
Denote by $\tilde{Z}_{v}$ the transformed version of the valid instrument
that appropriately takes into account the violations of the exclusion
and conditional exogeneity assumptions of the other instruments. Under
Assumption \ref{Ass:exoexcl} it follows that $\tilde{Z}_{v}\in\tilde{\boldsymbol{Z}}_{s_{v}}$
for an $s_{v}\in\left\{ 1,\ldots,S\right\} $. If the transformed
instrument $\tilde{Z}_{v}$ is relevant, $\tilde{\pi}_{v}\neq0$,
then it follows that $\tilde{\beta}_{v}=\frac{\tilde{\psi}_{v}}{\tilde{\pi}_{v}}=\text{\ensuremath{\beta}}$.
As $\tilde{\beta}_{v}\in FAS_{s_{v}}$, it follows that the generalized
$FAS$ is guaranteed to contain $\beta$ if there is a valid instrument
for which its correctly transformed version is relevant. We state
this result formally in the following proposition. 
\begin{prop}
\label{Prop:betainFAS}Under Assumptions \ref{Ass:Suff}, \ref{Ass:exoexcl}
and \ref{Ass:PartExclExo} consider the generalized FAS as defined
in Definition \ref{Def:GenFAS}. If there is at least one valid instrument
$Z_{\ell}$ with $\gamma_{\ell}=\alpha_{\ell}=0$ and this instrument
is relevant in the model specification that has included the invalid
instruments that violate the exclusion assumption in the model as
controls and excluded the invalid instruments that violate the conditional
exogeneity condition from the instrument set, then $\beta\in FAS_{G}$. 
\end{prop}

As an example, consider the case with $k_{z}=3$ and let $\boldsymbol{\gamma}+\boldsymbol{\alpha}=\left(0,\gamma_{2},\alpha_{3}\right)'$,
then $Z_{1|2}$ point-identifies $\beta$ if $\pi_{1|2}\neq0$. This
corresponds to the point-identified model specification $Y=X\beta+Z_{2}\gamma_{2}+\tilde{U}$,
$X=Z_{1}\pi_{1|2}+Z_{2}\pi_{2|1}+V$.

The generalized \textit{FAS} can become wide, as it is the union over all $2^{k_z}$ pattern-specific sets. However, it can be sharpened with a priori knowledge about violation types. For example, with $k_z = 3$ instruments, if the researcher knows that $Z_1$ can only violate exclusion, e.g. because it is not randomly assigned but has no plausible direct effect on the outcome, then only $2^2 = 4$ of the $2^3 = 8$ patterns need to be considered. If the exact pattern is known, the generalized \textit{FAS} reduces to a single pattern-specific set. In this way, knowledge about each instrument's potential invalidity directly sharpens the identified set.

The union structure across all $S=2^{k_z}$ patterns can equivalently be characterized through the collection of $J=k_{z}2^{k_{z}-1}$ distinct transformed instruments; see Appendix~\ref{app:TransInstUnion}. The sample analog $\hat{FAS}_G$ is given in Appendix~\ref{app:EstiGenFAS}, and the full computational procedure is summarized in Appendix~\ref{app:algorithm}.

\subsection{Endogenous Instruments}

\label{sec:Endogenous-Instruments}

If an invalid instrument $Z_{\ell}$ violates both the exclusion and
conditional exogeneity assumptions, so both $\gamma_{\ell}\neq0$
\textit{and} $\alpha_{\ell}\neq0$, then $Z_{\ell}$ itself an endogenous
explanatory variable. Whilst it is clear from the alternative derivation
of the falsification frontier in Section \ref{sec:Alternative-Derivation}
that each falsification frontier is a bounded set of relaxations that
results in nonfalsification, none of the associated falsification
adaptive sets, and hence the generalized \textit{FAS}, is now guaranteed
to contain $\beta$ even if there is a valid instrument present. That
is because an endogenous instrument needs itself to be instrumented
and the falsification frontiers are all based on just-identifying
linearly transformed instruments that represent inclusion of other
instruments as controls in the model and exclusion from the instrument
set. The restriction that $\gamma_{\ell}\alpha_{\ell}=0$ allows for
the counterfactual argument that if there is a valid instrument then
there is a \textit{FAS} that is an identified set for $\beta$.

\section{Empirical Example}

\label{sec:Empi}

We now revisit the DMT roads and trade application introduced in Section 1, which was also used by MP. We denote the three instruments as \textit{$Z_{1}=$ Plan} (1947 planned highways), \textit{$Z_{2}=$ Railroads} (1898 railroad routes), and \textit{$Z_{3}=$ Exploration} (historical exploration routes). We focus on the specification in column 2 of Table 5 in DMT (column 2 of Table I in MP), which includes ``log employment'' and ``Market access (export)'' as additional controls. Table \ref{tab:DurantonA} replicates the 2SLS results using all three instruments; this specification is falsified by the $J$-statistic (p-value $0.043$). The remaining columns give the twelve just-identified IV estimates from which the falsification adaptive sets are constructed.

\begin{table}[h]
\caption{ IV estimation results, \citet[Table 5, column 2]{Duranton2014}}\label{tab:DurantonA}

\begin{centering}
\begin{tabular}{cccccccc}
\hline 
 & \multicolumn{7}{c}{Instruments}\tabularnewline
\hline 
 & $Z_{1},Z_{2},Z_{3}$  & $Z_{1|2,3}$  & $Z_{2|1,3}$  & $Z_{3|1,2}$  & $Z_{1}$  & $Z_{2}$  & $Z_{3}$\tabularnewline
\hline 
log highway km  & $\begin{array}{c}
0.57\\
\ensuremath{\left(0.16\right)}
\end{array}$  & $\begin{array}{c}
0.28\\
\ensuremath{\left(0.25\right)}
\end{array}$  & $\begin{array}{c}
3.16\\
\ensuremath{\left(1.39\right)}
\end{array}$  & $\begin{array}{c}
-0.32\\
\ensuremath{\left(0.86\right)}
\end{array}$  & $\begin{array}{c}
0.55\\
\ensuremath{\left(0.17\right)}
\end{array}$  & $\begin{array}{c}
1.09\\
\ensuremath{\left(0.26\right)}
\end{array}$  & $\begin{array}{c}
0.13\\
\ensuremath{\left(0.38\right)}
\end{array}$\tabularnewline
$F$  & $90.30$  & $58.13$  & $6.97$  & $20.00$  & $154.5$  & $35.84$  & $15.97$\tabularnewline
$J$-test p-value  & $0.043$  &  &  &  &  &  & \tabularnewline
\hline 
 &  & $Z_{1|2}$  & $Z_{1|3}$  & $Z_{2|1}$  & $Z_{2|3}$  & $Z_{3|1}$  & $Z_{3|2}$\tabularnewline
\hline 
log highway km  &  & $\begin{array}{c}
0.22\\
\ensuremath{\left(0.21\right)}
\end{array}$  & $\begin{array}{c}
0.40\\
\ensuremath{\left(0.16\right)}
\end{array}$  & $\begin{array}{c}
3.74\\
\ensuremath{\left(1.90\right)}
\end{array}$  & $\begin{array}{c}
1.18\\
\ensuremath{\left(0.26\right)}
\end{array}$  & $\begin{array}{c}
-0.61\\
\ensuremath{\left(1.11\right)}
\end{array}$  & $\begin{array}{c}
-0.02\\
\ensuremath{\left(0.38\right)}
\end{array}$\tabularnewline
$F$  &  & $81.14$  & $122.45$  & $5.29$  & $34.31$  & $14.27$  & $31.07$\tabularnewline
\hline 
\end{tabular}
\par\end{centering}
{\small Notes: Outcome variable ``propensity to export weight'',
$n=66$. Additional controls ``log employment'' and ``Market access
(export)''. Heteroskedasticity robust test statistics and (standard
errors). $Z_{1}$ is instrument ``Plan'', $Z_{2}$ is ``Railroads'',
$Z_{3}$ is ``Exploration''. $Z_{\ell|B}$ is the just-identifying
instrument $Z_{\ell}$ in the model with $\boldsymbol{Z}_{B}$ included
as controls.}{\small\par}
\end{table}

\begin{table}[h]
\caption{Falsification Adaptive Sets}\label{tab:Fas}

\begin{centering}
\begin{tabular}{lll@{\hskip 1.5em}c@{\hskip 1.5em}ll}
\toprule
 & & \multicolumn{2}{c}{All relevant} & \multicolumn{2}{c}{$F\geq10$} \tabularnewline
\cmidrule(lr){3-4} \cmidrule(lr){5-6}
$\left(\boldsymbol{\gamma}_{s}+\boldsymbol{\alpha}_{s}\right)'$ & $\tilde{\boldsymbol{Z}}_{s,\ell}^{\prime}$ & $\hat{FAS}{}_{s}$ & $\hat{FAS}_{G}$ & $\hat{FAS}{}_{s}^{F\geq10}$ & $\hat{FAS}_{G}{}^{F\geq10}$\tabularnewline
\midrule
$\alpha_{1}\,\alpha_{2}\,\alpha_{3}$ & $Z_{1}\,Z_{2}\,Z_{3}$ & $\left[0.13,1.09\right]$ & $\left[-0.61,3.74\right]$ & $\left[0.13,1.09\right]$ & $\left[-0.61,1.18\right]$\tabularnewline[3pt]
$\alpha_{1}\,\alpha_{2}\,\gamma_{3}$ & $Z_{1|3}\,Z_{2|3}\,Z_{3}$ & $\left[0.13,1.18\right]$ &  & $\left[0.13,1.18\right]$ & \tabularnewline[3pt]
$\alpha_{1}\,\gamma_{2}\,\alpha_{3}$ & $Z_{1|2}\,Z_{2}\,Z_{3|2}$ & $\left[-0.02,1.09\right]$ &  & $\left[-0.02,1.09\right]$ & \tabularnewline[3pt]
$\gamma_{1}\,\alpha_{2}\,\alpha_{3}$ & $Z_{1}\,Z_{2|1}\,Z_{3|1}$ & $\left[-0.61,3.74\right]$ &  & $\left[-0.61,0.55\right]$ & \tabularnewline[3pt]
$\alpha_{1}\,\gamma_{2}\,\gamma_{3}$ & $Z_{1|23}\,Z_{2|3}\,Z_{3|2}$ & $\left[-0.02,1.18\right]$ &  & $\left[-0.02,1.18\right]$ & \tabularnewline[3pt]
$\gamma_{1}\,\alpha_{2}\,\gamma_{3}$ & $Z_{1|3}\,Z_{2|13}\,Z_{3|1}$ & $\left[-0.61,3.16\right]$ &  & $\left[-0.61,0.40\right]$ & \tabularnewline[3pt]
$\gamma_{1}\,\gamma_{2}\,\alpha_{3}$ & $Z_{1|2}\,Z_{2|1}\,Z_{3|12}$ & $\left[-0.32,3.74\right]$ &  & $\left[-0.32,0.22\right]$ & \tabularnewline[3pt]
$\gamma_{1}\,\gamma_{2}\,\gamma_{3}$ & $Z_{1|23}\,Z_{2|13}\,Z_{3|12}$ & $\left[-0.32,3.16\right]$ &  & $\left[-0.32,0.28\right]$ & \tabularnewline
\bottomrule
\end{tabular}
\par\end{centering}
\end{table}

Table \ref{tab:Fas} presents the estimated falsification adaptive sets for all eight patterns of violations. MP propose to deem an instrument irrelevant if its $F$-statistic is less than $10$; the instruments $Z_{2|13}$ and $Z_{2|1}$ are classified as irrelevant under this cutoff (p-values $0.0106$ and $0.0249$), though with $n=66$ one could argue they are relevant. Treating all instruments as relevant, $\hat{FAS}_{exo}=\left[0.13,1.09\right]$ is the smallest set and $\hat{FAS}_{excl}=\left[-0.32,3.16\right]$. The generalized $\hat{FAS}_{G}=\left[-0.61,3.74\right]$ is a wide interval reflecting all possible violation patterns. Its lower bound is driven by the IV estimate of $Z_{3|1}$, which is the valid instrument when $\boldsymbol{\gamma}+\boldsymbol{\alpha}=\left(\gamma_{1},\alpha_{2},0\right)'$; its upper bound by $Z_{2|1}$, valid when $\boldsymbol{\gamma}+\boldsymbol{\alpha}=\left(\gamma_{1},0,\alpha_{3}\right)'$.

One might suspect that $\hat{FAS}_G$ is wide because it is the union of all eight pattern-specific sets. However, its bounds coincide with those of a single $\hat{FAS}_s$ for $\boldsymbol{\gamma}_s+\boldsymbol{\alpha}_s = \left(\gamma_1,\alpha_2,\alpha_3\right)'$, so the width is not an artefact of the union but reflects a violation pattern that cannot be excluded. Nor is it driven by weak instruments: even after prescreening with $F\geq 10$, zero is contained in $\hat{FAS}_s^{F\geq 10}$ in all but two pattern-specific sets.

The availability of narrower pattern-specific sets in Table \ref{tab:Fas} might tempt the researcher to simply select one, such as $\hat{FAS}_{exo}$ or $\hat{FAS}_{excl}$. However, doing so amounts to cherry-picking. As shown in Section \ref{sec:ExoExcl}, a priori knowledge about violation types can legitimately restrict the set of relevant patterns and sharpen the \textit{FAS}, but this must be decided beforehand and motivated by theoretical or institutional knowledge.

Note again that the 2SLS estimator and the test for overidentifying
restrictions are the same for all eight linear transformations of
the instruments. The 2SLS estimator is a linear combination of the
just-identifying estimates. The weights for the exogeneity and exclusion
only specifications, $\left\{ Z_{1},Z_{2},Z_{3}\right\} $ and $\left\{ Z_{1|2,3},Z_{2|1,3},Z_{3|1,2}\right\} $
are identical\footnote{See \citet[pp 14-15]{Apfel2022}.} and here
given by $w=\left\{ 0.757,0.126,0.117\right\} $. So whilst $Z_{2|1,3}$
is found to be irrelevant using the MP cutoff for the $F$-statistic,
as described below, its estimate of $\beta$ contributes to the 2SLS
estimate with a substantial weight, larger than that of $Z_{3|1,2}$
which is found to be relevant. It is the same weight as for the
$Z_{2}$ estimate of $\beta$ in the exogeneity only specification,
which is found to be relevant also with the MP cutoff. The overidentification
test is testing whether the three just-identifying IV estimators estimate
the same value for $\beta$, see \citet{Windmeijer2019}. As the test
rejects, it does seem appropriate to consider all just-identified
estimates for the construction of each falsification adaptive set,
and not to differentially omit the estimates from just-identifying
instruments that are classified as irrelevant, as it changes the specification,
as we illustrate next.

When using MP's $F$-statistic cutoff of $10$, the instruments $Z_{2|1,3}$ and $Z_{2|1}$ are classified as irrelevant. Since these produce the two largest IV estimates ($3.16$ and $3.74$), removing them substantially narrows the pattern-specific sets (column 5 of Table \ref{tab:Fas}). For example, $\hat{FAS}_{excl}^{F\geq10}=\left[-0.32,0.28\right]$. However, omitting $Z_{2|13}$ from $\hat{FAS}_{excl}$ implicitly changes the model specification to $$Y=X\beta+Z_2\gamma_2+U$$ with $\left\{Z_1,Z_3\right\}$ as instruments. This specification is not falsified ($J$-test p-value $0.428$). The \textit{FAS} is then no longer motivated by falsification of the baseline model.

\section{Conclusions}

\label{sec:Conclusions}

In this paper, we have studied settings with multiple potentially invalid IVs and shown that different types of violations have important consequences for the falsification adaptive set of \citet{MastenPoirierEcta2021}. Some invalid instruments may violate the exclusion restriction, calling for the treatment of these instruments as confounders, while others may violate exogeneity, calling for the treatment of these as colliders. This distinction affects the \textit{FAS}: each pattern of violations across instruments gives rise to a different pattern-specific \textit{FAS}. We propose the generalized \textit{FAS} as the union of all pattern-specific sets, which allows the researcher to remain agnostic about the pattern of violations. If at least one instrument is valid and relevant, the generalized \textit{FAS} is guaranteed to contain $\beta$, and we recommend researchers to report estimates of this set when their baseline model is falsified.

\bibliographystyle{plainnat}
\bibliography{false}

\newpage{}

\section*{Appendix }

\global\long\def\thesection{A}%
\global\long\def\theequation{A.\arabic{equation}}%
\global\long\def\thetable{A\arabic{table}}%
\setcounter{table}{0}\setcounter{equation}{0}

\subsection{Proof Sketch for Proposition \ref{Prop:IdSetExoExcl}}

\label{app:PropIdSetExoExcl} The proof follows the same arguments
as those of Theorem 1 in MP. For each $s\in\left\{ 1,\ldots,S\right\} $,
it is clear that any value $\beta$ consistent with the model lies
in $\mathcal{B}_{s}\left(\tilde{\boldsymbol{\omega}}_{s}\right)$.
For any $b\in\mathcal{B}_{s}\left(\tilde{\boldsymbol{\omega}}_{s}\right)$
define 
\[
\boldsymbol{\gamma}_{s}\left(b\right)+\tilde{\boldsymbol{\alpha}}_{s}\left(b\right)\coloneqq\text{cov}\left(\boldsymbol{Z}_{s}^{*},Y-Xb\right),
\]
then $-\tilde{\boldsymbol{\omega}}_{s}\leq\boldsymbol{\gamma}_{s}\left(b\right)+\tilde{\boldsymbol{\alpha}}_{s}\left(b\right)\leq\tilde{\boldsymbol{\omega}}_{s}$.
If we specify for $\ell\in\mathcal{C}_{s}$, 
\[
\gamma_{s,\ell}\left(b\right):=\text{cov}\left(Z_{s,\ell}^{*},Y-Xb\right),
\]
then $\tilde{\alpha}_{s,l}\left(b\right)=0$. If we specify, for $\ell\in\mathcal{A}_{s}$,
\[
\tilde{\alpha}_{s,\ell}\left(b\right):=\text{cov}\left(Z_{s,\ell}^{*},Y-Xb\right),
\]
then $\gamma_{s,\ell}\left(b\right)=0$, hence $\mathcal{B}_{s}\left(\tilde{\boldsymbol{\omega}}_{s}\right)$
is sharp.

\subsection{Estimation of \textit{FAS}$_{excl}$ and First-Stage Hard Thresholding}

\label{app:EstiFSHT}

We have an i.i.d. sample of size $n$, $\left\{ Y_{i},X_{i},\boldsymbol{Z}_{i}^{\prime}\right\} _{i=1}^{n}$.
The $n$-vectors $\left(Y_{i}\right)$ and $\left(X_{i}\right)$ are
denoted $\boldsymbol{y}$ and \textbf{$\boldsymbol{x}$} respectively,
and here $\boldsymbol{Z}$ denotes the $n\times k_{z}$ matrix of
observations on the instruments. The constant and other exogenous
variables have been partialled out. MP suggest to estimate the set
of relevant instruments by 
\[
\mathcal{\hat{L}}_{rel}=\left\{ \ell\in\left\{ 1,\ldots,k_{z}\right\} :F_{\ell}\geq C_{n}\right\} ,
\]
where $F_{\ell}$ is the first-stage $F$-statistic for model (\ref{eq:jidcon}),
where $Z_{\ell}$ is considered as an instrument and $Z_{\left\{ -\ell\right\} }$
as controls. For all values of $\ell$, the first-stage model is therefore
given by $\boldsymbol{x}=\boldsymbol{Z}\boldsymbol{\pi}+\boldsymbol{v}$
and so $F_{\ell}$ is the same as the Wald statistic for testing $H_{0}:\pi_{\ell}=0$
based on the OLS estimator of $\boldsymbol{\pi}$, denoted $\hat{\boldsymbol{\pi}}$.
The same first-stage hard thresholding was proposed in \citet{GuoetalJRSSB2018}.
Although $C_{n}\rightarrow\infty$ as $n\rightarrow\infty$ and $C_{n}=o\left(n\right)$
for consistent selection, MP choose $C_{n}=10$ as their default cutoff,
or for the t-ratio, $\left|\frac{\hat{\pi}_{\ell}}{se\left(\hat{\pi}_{\ell}\right)}\right|\geq\sqrt{10}=3.16$.

Let $\hat{\beta}_{\ell}$ be the IV estimator of $\beta_{\ell}$ in
just-identified model specification (\ref{eq:jidcon}). Then \textit{FAS$_{excl}$}
is estimated by 
\[
\hat{FAS}_{excl}=\left[\min_{\ell\in\mathcal{\hat{L}}_{rel}}\hat{\beta}_{\ell},\max_{\ell\in\mathcal{\hat{L}}_{rel}}\hat{\beta}_{\ell}\right]
\]
and MP show that $\hat{FAS}_{excl}$ is a consistent estimator of
the $FAS_{excl}$ under the conditions of their Proposition 3.

\subsection{Estimation of \textit{FAS}$_{exo}$}

\label{app:EstiFASexo}

The set of relevant instruments can here be estimated by 
\[
\mathcal{\hat{L}}_{rel}^{*}=\left\{ \ell\in\left\{ 1,\ldots,k_{z}\right\} :F_{\ell}^{*}\geq C_{n}\right\} ,
\]
where $F_{\ell}^{*}$ is the F-statistic for testing $H_{0}:\pi_{\ell}^{*}=0$
in the first-stage linear specification $\boldsymbol{x}=\boldsymbol{z}_{\ell}\pi_{\ell}^{*}+\boldsymbol{v}_{\ell}.$
Let $\hat{\beta}_{\ell}^{*}=\frac{\boldsymbol{z}_{\ell}^{\prime}\boldsymbol{y}}{\boldsymbol{z}_{\ell}^{\prime}\boldsymbol{x}}$,
where again the constant and other exogenous variables have been partialled
out, then the consistent estimator of \textit{FAS$_{exo}$} is given
by 
\[
\hat{FAS}_{exo}=\left[\min_{\ell\in\mathcal{\hat{L}}_{rel}^{*}}\hat{\beta}_{\ell}^{*},\max_{\ell\in\mathcal{\hat{L}}_{rel}^{*}}\hat{\beta}_{\ell}^{*}\right].
\]

\subsection{Transformed Instruments: Union Across Patterns}

\label{app:TransInstUnion}

The pattern-specific $\tilde{\boldsymbol{Z}}_{s}$ constructed in Section~\ref{sec:ExoExcl} overlap across $s$: the same partialled-out instrument can appear in several $\tilde{\boldsymbol{Z}}_{s}$. It is therefore useful to reindex the union of the pattern-specific instruments by a single subscript $j=1,\ldots,J$.

Although there are $S=2^{k_{z}}$ different patterns $\boldsymbol{\gamma}_{s}+\boldsymbol{\alpha}_{s}$,
it is clear that there is overlap of the linearly transformed instruments
when constructing the different \textit{FAS$_{s}$}. There are a total
of $J=k_{z}\,2^{k_{z}-1}$ transformed instruments. Let $\tilde{\boldsymbol{Z}}$
denote the $J$-vector of transformed instruments. For example, for
$k_{z}=3$, we have the following set of $J=12$ transformed instruments
\[
\tilde{\boldsymbol{Z}}=\left(Z_{1},Z_{1|2},Z_{1|3},Z_{1|23},Z_{2},Z_{2|1},Z_{2|3},Z_{2|13},Z_{3},Z_{3|1},Z_{3|2},Z_{3|12}\right)'.
\]

\noindent Then define the $J$-vectors $\tilde{\boldsymbol{\pi}}$
and $\tilde{\boldsymbol{\psi}}$ with elements 
\begin{equation}
\tilde{\pi}_{j}=\left(\text{var}\left(\tilde{Z}_{j}\right)\right)^{-1}\text{cov}\left(\tilde{Z}_{j},X\right);\,\,\tilde{\psi}_{j}=\left(\text{var}\left(\tilde{Z}_{j}\right)\right)^{-1}\text{cov}\left(\tilde{Z}_{j},Y\right),\label{eq:pitjpsitj}
\end{equation}
 for $j=1,\ldots,J$. Let
\[
\mathcal{\tilde{\mathcal{L}}}_{rel}=\left\{ j\in\left\{ 1,\ldots,J\right\} :\tilde{\pi}_{j}\neq0\right\} .
\]

When all transformed instruments are relevant, so $\mathcal{\tilde{\mathcal{L}}}_{rel}=\left\{ 1,\ldots,J\right\} $,
then it is straightforward to show that
\[
FAS_G=\left[\min_{j}\frac{\tilde{\psi}_{j}}{\tilde{\pi}_{j}},\max_{j}\frac{\tilde{\psi}_{j}}{\tilde{\pi}_{j}}\right],
\]
as there are then always overlapping pattern-specific \textit{FAS}$_{s}$,
from the one containing $\min_{j}\frac{\tilde{\psi}_{j}}{\tilde{\pi}_{j}}$
to the one containing $\max_{j}\frac{\tilde{\psi}_{j}}{\tilde{\pi}_{j}}$.
When not all transformed instruments are relevant, the generalized
\textit{FAS} could be a set of disjoint intervals. However, it is
trivially the case that
\[
FAS_G=\left[\min_{j\in\mathcal{\tilde{\mathcal{L}}}_{rel}}\frac{\tilde{\psi}_{j}}{\tilde{\pi}_{j}},\max_{j\in\mathcal{\tilde{\mathcal{L}}}_{rel}}\frac{\tilde{\psi}_{j}}{\tilde{\pi}_{j}}\right],
\]
if there is a set of overlapping \textit{FAS}$_{s}$ that contain
$\min_{j\in\mathcal{\tilde{\mathcal{L}}}_{rel}}\frac{\tilde{\psi}_{j}}{\tilde{\pi}_{j}}$
and $\max_{j\in\mathcal{\tilde{\mathcal{L}}}_{rel}}\frac{\tilde{\psi}_{j}}{\tilde{\pi}_{j}}$.
We will illustrate this with estimated values in the empirical application
in Section \ref{sec:Empi}.

\subsection{Estimation of the Generalized \textit{FAS}}

\label{app:EstiGenFAS}

We have again partialled out the constant and other exogenous variables.
For the pattern-specific falsification adaptive sets, FAS$_{s}$,
for each $s\in$$\left\{ 1,\ldots,S\right\} $, with $S=2^{k_{z}}$,
for each $\ell\in\mathcal{C}_{s}$ the transformation of the $n$-vector
of observations $\boldsymbol{z}_{\ell}$ is given by
\[
\tilde{\boldsymbol{z}}_{s,\ell}=\boldsymbol{M}_{Z_{\mathcal{C}_{s,-\ell}}}\boldsymbol{z}_{\ell},
\]
where for a general full column rank matrix $\boldsymbol{A}$, $\boldsymbol{M}_{A}=\boldsymbol{I}_{n}-\boldsymbol{A}\left(\boldsymbol{A}'\boldsymbol{A}\right)^{-1}\boldsymbol{A}'$,
with $\boldsymbol{I}_{n}$ is the $n$-dimensional identity matrix.
For each $\ell\in\mathcal{A}_{s}$ the transformation of the $n$-vector
of observations $\boldsymbol{z}_{\ell}$ is given by
\[
\tilde{\boldsymbol{z}}_{s,\ell}=\boldsymbol{M}_{Z_{\mathcal{C}_{s}}}\boldsymbol{z}_{\ell}.
\]
The $n\times k_{z}$ matrix of pattern-specific transformed instruments
is then given by
\[
\tilde{\boldsymbol{Z}}_{s}=\left[\tilde{\boldsymbol{z}}_{s,\ell}\right].
\]

For each $s$, the set of relevant instruments is estimated by

\[
\mathcal{\hat{\tilde{L}}}_{s,rel}=\left\{ \ell\in\left\{ 1,\ldots,k_{z}\right\} :\tilde{F}_{s,\ell}\geq C_{n}\right\} ,
\]
where $\tilde{F}_{s,\ell}$ is the F-statistic for testing $H_{0}:\tilde{\pi}_{s,\ell}=0$
in the first-stage linear specification $\boldsymbol{x}=\tilde{\boldsymbol{z}}_{s,\ell}\tilde{\pi}_{s,\ell}+\boldsymbol{v}_{s,\ell}.$
Let $\hat{\tilde{\beta}}_{s,\ell}=\frac{\tilde{\boldsymbol{z}}_{s,\ell}^{\prime}\boldsymbol{y}}{\tilde{\boldsymbol{z}}_{s,\ell}^{\prime}\boldsymbol{x}}$,
then the consistent estimator of \textit{FAS$_{s}$} is given by 
\[
\hat{FAS}_{s}=\left[\min_{\ell\in\mathcal{\hat{\tilde{L}}}_{s,rel}}\hat{\tilde{\beta}}_{s,\ell},\max_{\ell\in\mathcal{\hat{\tilde{L}}}_{s,rel}}\hat{\tilde{\beta}}_{s,\ell}\right].
\]
The consistent estimator of the generalized falsification adaptive
set is then obtained as
\[
\hat{FAS}_G=\bigcup_{s=1}^{S}\hat{FAS}_{s}.
\]

Let the $n\times J$ matrix of transformed instruments be 
\[
\tilde{\boldsymbol{Z}}=\cup_{s=1}^{S}\tilde{\boldsymbol{Z}}_{s},
\]
with $J=k_{z}2^{k_{z}-1}$. Then let
\[
\mathcal{\hat{\tilde{L}}}_{rel}=\left\{ j\in\left\{ 1,\ldots,J\right\} :\tilde{F}_{j}\geq C_{n}\right\} ,
\]
where $\tilde{F}_{j}$ is the F-statistic for testing $H_{0}:\tilde{\pi}_{j}=0$
in the first-stage linear specification $\boldsymbol{x}=\tilde{\boldsymbol{z}}_{j}\tilde{\pi}_{j}+\boldsymbol{v}_{j}$. 

Let $\hat{\tilde{\beta}}_{j}=\frac{\tilde{\boldsymbol{z}}_{j}^{\prime}\boldsymbol{y}}{\tilde{\boldsymbol{z}}_{j}^{\prime}\boldsymbol{x}}$.
If all transformed instruments are relevant, so $\mathcal{\hat{\tilde{L}}}_{rel}=\left\{ 1,\ldots,J\right\} $,
then
\[
\hat{FAS}_G=\left[\min_{j\in\left\{ 1,\ldots,J\right\} }\hat{\tilde{\beta}}_{j},\max_{j\in\left\{ 1,\ldots,J\right\} }\hat{\tilde{\beta}}_{j}\right].
\]
When not all transformed instruments are relevant, it is the case
that 
\[
\hat{FAS}_G=\left[\min_{j\in\mathcal{\hat{\tilde{L}}}_{rel}}\hat{\tilde{\beta}}_{j},\max_{j\in\mathcal{\hat{\tilde{L}}}_{rel}}\hat{\tilde{\beta}}_{j}\right]
\]
if there can be found a path of overlapping $\hat{FAS}_{s}$ intervals
from the $\hat{FAS}_{s}$ that contains $\min_{j\in\mathcal{\hat{\tilde{L}}}_{rel}}\hat{\tilde{\beta}}_{j}$
to the $\hat{FAS}_{s}$ that contains $\max_{j\in\mathcal{\hat{\tilde{L}}}_{rel}}\hat{\tilde{\beta}}_{j}$,
as we demonstrate in the empirical example in Section \ref{sec:Empi}.

\subsection{Computing the Generalized FAS}

\label{app:algorithm}

The following pseudocode describes the computation of the generalized \textit{FAS} or any restriction thereof.

\begin{enumerate}
\item[\textbf{Step 0.}] \textbf{Choose the set of violation patterns.}
\begin{enumerate}
\item[(i)] If the researcher is fully agnostic, consider all $S = 2^{k_z}$ patterns.
\item[(ii)] If the researcher has a priori knowledge that some instruments can only violate a specific assumption, restrict to the $S' < S$ patterns consistent with this knowledge.
\item[(iii)] If the researcher is certain about the exact violation pattern, set $S' = 1$.
\end{enumerate}

\item[\textbf{Step 1.}] \textbf{Compute sample moments.} From the data $\{Y_i, X_i, \boldsymbol{Z}_i'\}_{i=1}^n$, compute $\widehat{\text{var}}(\boldsymbol{Z})$, $\widehat{\text{cov}}(\boldsymbol{Z}, X)$, and $\widehat{\text{cov}}(\boldsymbol{Z}, Y)$.

\item[\textbf{Step 2.}] \textbf{For each pattern $s = 1, \ldots, S'$:}
\begin{enumerate}
\item[(i)] Determine $\mathcal{C}_s$ (instruments treated as confounders, violating exclusion) and $\mathcal{A}_s = \mathcal{K} \setminus \mathcal{C}_s$ (instruments treated as colliders, violating exogeneity).
\item[(ii)] For each $\ell \in \mathcal{C}_s$: partial out $\boldsymbol{Z}_{\mathcal{C}_{s,-\ell}}$ from $Z_\ell$ to obtain the transformed instrument $\tilde{Z}_{s,\ell} = Z_{\ell|\mathcal{C}_{s,\{-\ell\}}}$.
\item[(iii)] For each $\ell \in \mathcal{A}_s$: partial out $\boldsymbol{Z}_{\mathcal{C}_s}$ from $Z_\ell$ to obtain $\tilde{Z}_{s,\ell} = Z_{\ell|\mathcal{C}_s}$.
\item[(iv)] Compute the just-identified IV estimates: $\hat{\tilde{\beta}}_{s,\ell} = \widehat{\text{cov}}(\tilde{Z}_{s,\ell}, Y) \,/\, \widehat{\text{cov}}(\tilde{Z}_{s,\ell}, X)$.
\item[(v)] Determine the set of relevant instruments $\tilde{\mathcal{L}}_{s,rel}$ (e.g., using an $F$-statistic cutoff).
\item[(vi)] Compute $\hat{FAS}_s = \left[\min_{\ell \in \tilde{\mathcal{L}}_{s,rel}} \hat{\tilde{\beta}}_{s,\ell},\; \max_{\ell \in \tilde{\mathcal{L}}_{s,rel}} \hat{\tilde{\beta}}_{s,\ell}\right]$.
\end{enumerate}

\item[\textbf{Step 3.}] \textbf{Take the union.} $\hat{FAS}_G = \bigcup_{s=1}^{S'} \hat{FAS}_s$.
\end{enumerate}

\noindent When $S' = S$, Step 3 yields the fully agnostic generalized \textit{FAS}. When $S' = 1$, it yields a single pattern-specific \textit{FAS}, such as $\hat{FAS}_{excl}$ or $\hat{FAS}_{exo}$.

\end{document}